\documentclass[aps,twocolumn,showpacs,superscriptaddress]{revtex4}

\usepackage{graphicx}
\usepackage{bm}
\usepackage{amsmath}
\usepackage{dcolumn}%
\usepackage{epstopdf}
\usepackage{csquotes}
\usepackage{xcolor}

\newcommand{\eg}{\ensuremath{e_{g}}}
\newcommand{\tg}{\ensuremath{t_{2g}}}

\newcommand{\mb}{\ensuremath{\mu_{\text{B}}}}

\newcolumntype{/}{D{/}{/}{2,2}}  
\newcolumntype{.}{D{.}{.}{0}}  

\begin{document}

\title{\emph{Ab initio} modeling of resonant inelastic x-ray scattering from Ca$_2$RuO$_4$}

\author{D. A. Kukusta}

\affiliation{G. V. Kurdyumov Institute for Metal Physics of the
  N.A.S. of Ukraine, 36 Academician Vernadsky Boulevard, UA-03142
  Kyiv, Ukraine}

\author{L. V. Bekenov}

\affiliation{G. V. Kurdyumov Institute for Metal Physics of the
  N.A.S. of Ukraine, 36 Academician Vernadsky Boulevard, UA-03142
  Kyiv, Ukraine}

\author{P. F. Perndorfer}
\affiliation{Institute for Theoretical Physics, Johannes Kepler University, Altenberger Strasse 69, A-4040 Linz, Austria}
\affiliation{Department of Engineering and Computer Sciences, Hamburg University of Applied Sciences, Berliner Tor 7, D-20099 Hamburg, Germany}

\author{D.\,V. Vyalikh}
\affiliation{Donostia International Physics Center (DIPC), 20018 Donostia-San Sebasti\'{a}n, Spain}
\affiliation{IKERBASQUE, Basque Foundation for Science, 48011 Bilbao, Spain}

\author{P. A. Buczek}
\affiliation{Department of Engineering and Computer Sciences, Hamburg University of Applied 
             Sciences, Berliner Tor 7, D-20099 Hamburg, Germany}            

\author{A. Ernst}

\affiliation{Institute for Theoretical Physics, Johannes Kepler University, Altenberger Strasse 69, A-4040 Linz, Austria}

\affiliation{Max Planck Institute of Microstructure Physics, D-06120 Halle
  (Saale), Germany }

\author{V. N. Antonov}

\affiliation{G. V. Kurdyumov Institute for Metal Physics of the
  N.A.S. of Ukraine, 36 Academician Vernadsky Boulevard, UA-03142
  Kyiv, Ukraine}

\affiliation{Max-Planck-Institute for Solid State Research,
  Heisenbergstrasse 1, 70569 Stuttgart, Germany}

\date{\today}

\begin{abstract}

The single-layered perovskite Ca$_2$RuO$_4$, characterized by a 4$d^4$ electron configuration, 
has been studied from first principles using density functional theory (DFT) using the generalized 
gradient approximation, with inclusion of strong on-site Coulomb interactions and spin-orbit 
coupling (GGA+SO+$U$), in the framework of the fully relativistic, spin-polarized Dirac linear 
muffin-tin orbital (LMTO) band-structure method. This approach enabled a comprehensive investigation 
of the electronic structure of Ca$_2$RuO$_4$ through the modeling of relevant spectra obtained 
from synchrotron-based techniques widely used to probe electronic properties, with a primary focus on
resonant inelastic X-ray scattering (RIXS) at the Ru $L_3$ and O $K$ edges. The calculated spectra 
were thoroughly analyzed with available experimental data reported in the literature. The good 
agreement between our results and experimental observations for Ca$_2$RuO$_4$ enables a conclusive 
interpretation of key features in the spectra obtained from the aforementioned techniques. 
Consequently, this enables us to describe its electronic properties and to establish a solid 
theoretical approach suitable for routine modeling of spectra, particularly from RIXS, aimed at 
characterizing the electronic structure and properties of similar or more complex strongly 
correlated, technologically relevant materials.

\end{abstract}


\maketitle

\section{Introduction}

\label{sec:introd}

Due to the extended nature of 4$d$ and 5$d$ wave functions and the resulting broad energy bands in 
solids, 4$d$ and 5$d$ transition-metal oxides are naturally expected to behave as weakly correlated 
materials. However, driven by strong spin-orbit coupling (SOC), many of these compounds exhibit 
a Mott-insulating state and, as a consequence, display unusual electronic and magnetic 
properties~\cite{KJM+08,KOK+09}. The strong SOC in $d$ systems split the {\tg} orbitals into a 
Kramers quartet ($J_{\rm{eff}}$ = 3/2) and a doublet ($J_{\rm{eff}}$ = 1/2)
\cite{JaKh09,CPB10,WCK+14}. When, in addition to strong SOC, electron correlations emerge, this can 
give rise to such fascinating phenomena as Mott insulators~\cite{KJM+08,KOK+09,JaKh09,WSY10,MAV+11}, 
topological insulators~\cite{QZ10,Ando13,WBB14,BLD16}, Weyl semimetals ~\cite{WiKi12,GWJ12,SHJ+15}, 
and quantum spin liquids~\cite{JaKh09,KAV14}. So far, most research efforts have focused on $d^{4+}$ 
systems with a $t_{2g}^5$ configuration~\cite{KJM+08,VEG+17,HAE+18,ABK20,AKU+21}. For such materials, 
for example, the layered perovskite Sr$_2$IrO$_4$, the quartet $J_{\rm{eff}}$ = 3/2 is fully occupied, 
and the relatively narrow $J_{\rm{eff}}$ = 1/2 doublet occupied by one electron can be already split 
by moderate Hubbard $U_{\rm{eff}}$ with opening a small band gap called the relativistic
Mott gap~\cite{KJM+08,MAV+11,AUU18}.

Another broad family of magnetic Mott-insulating $d$-electron systems, which often exhibit 
peculiar electronic and magnetic properties, is that with a completely empty Kramers upper 
$J_{\rm{eff}} = 1/2$ doublet and a partially filled lower $J_{\rm{eff}} = 3/2$ quartet of the 
{\tg} orbitals. There are several possibilities with partially filled
{\tg} orbitals, namely, $d^1$, $d^2$, $d^3$, and $d^4$ cases. Mott
insulators with $d^1$ and $d^2$ configurations have been shown to
exhibit exotic magnetic phases~\cite{DCK11,ChBa11,MOR+13,SZR+21} in
the presence of large SOC. In the $d^3$ case, SOC is quenched in a
cubic environment~\cite{MOR+13}. For a $d^4$ electron configuration
({\tg}$^4${\eg}$^0$) the system is expected to be nonmagnetic in both
the weakly and strongly correlated limits. In the weakly correlated
picture, when the SOC dominates over the Hund's coupling, {\tg} shells
are split into a fully filled $J_{\rm{eff}}$ = 3/2 shell and an empty
$J_{\rm{eff}}$ = 1/2 shell due to strong SOC. SOC then opens up a band
gap between the $J_{\rm{eff}}$ = 3/2 and $J_{\rm{eff}}$ = 1/2 bands
leading to a nonmagnetic insulating ground state. In the strongly
correlated picture, the first two Hund's rules require each $d^4$ site
to have total $S$ = 1 and $L$ = 1. The SOC then yields a local $J$
= 0 state on every ion with a nonmagnetic ground state~\cite{ChBa11,CSZ+17}.

Although octahedrally coordinated Re$^{3+}$, Os$^{4+}$, Ir$^{5+}$, and Ru$^{4+}$ systems 
with a $d^4$ electronic configuration have been known since the 1960s ~\cite{EFL+61}, 
they have been largely overlooked over the years. However, following the report of long-range 
magnetic order in Sr$_2$YIrO$_6$~\cite{CQL+14} and the proposal of a condensation mechanism 
for magnetism in $d^4$ Mott insulators~\cite{Kha13}, $d^4$ electron systems have recently attracted 
significant interest.

We consider here the electronic structure and physical properties of the single-layered perovskite 
Ca$_2$RuO$_4$ which belongs to the so-called Ruddlesden-Popper (RP) type ruthenates 
A$_{n+1}$Ru$_n$O$_{3n+1}$ (A = Sr or Ca), where $n$ is the number of Ru-O layers per unit cell. These 
ruthenates have recently received significant attention due to the discovery of their fascinating 
physical properties. For example, Sr$_2$RuO$_4$ ($n = 1$) exhibits unconventional $p$-wave 
superconductivity~\cite{MHY+94,MRS01}, whereas Ca$_2$RuO$_4$ is a typical Mott insulator~\cite{CMS+97}. 
In the $n$ = $\infty$ family SrRuO$_3$ is a ferromagnetic metal ~\cite{CMS+97}, whereas CaRuO$_3$ 
does not show any magnetic ordering~\cite{KAG+99}.

The Ca$_2$RuO$_4$ discussed here is a topical material exhibiting a wealth of physical properties. 
Among them, a record-high nonsuperconducting diamagnetic response has been reported~\cite{SYK+17}. 
Superconductivity emerges in strained films~\cite{cm:NYK+17} or upon application of hydrostatic 
pressure to bulk crystals~\cite{ANG+10}. Next, Ca$_2$RuO$_4$ is reported to undergo a series of 
phase transitions upon cooling: a metal-to-insulator transition at 357~K ~\cite{BAN+98,ACD+99,FBA+01,SFA+05}, 
orbital ordering at 260~K~\cite{ZSN+05,PGF+18}, and antiferromagnetic (AFM) ordering at 
110~K~\cite{BAN+98,ACD+99,ZSN+05,PGF+18}. Neutron and Raman scattering experiments have demonstrated 
the existence of a spin-orbit exciton ~\cite{KKS+15,JKP+17,SCK+17,GSK+19}. Because it remains 
insulating above the N$\rm{\acute{e}}$el temperature and exhibits Curie-Weiss magnetic susceptibility, 
Ca$_2$RuO$_4$~\cite{NIM97} is widely recognized as a Mott insulator. The momentum-resolved photoemission 
(ARPES) measurements of the paramagnetic insulating band structure~\cite{SFM+17} have also been 
interpreted in favor of an orbitally differentiated band-Mott insulating ground state~\cite{GKW+10}. 
This rich phenomenology of Ca$_2$RuO$_4$ reflects the intricate interplay between the Coulomb 
interaction $U$, Hund's coupling $J_H$, crystal electric field (CEF) splitting, and SOC.

The essential part of our work is dedicated to a comprehensive $ab-initio$ theoretical 
characterization of the electronic structure of Ca$_2$RuO$_4$, with a focus on modeling of Resonant 
Inelastic X-ray Scattering (RIXS) spectra at the Ru $L_3$ and O $K$ edges of this material, along 
with their analysis, interpretation, and comparison with existing literature data. Since the first 
publication by Kao {\it et al.} on NiO
~\cite{KCH+96}, the RIXS method has shown remarkable progress as a
spectroscopic technique to record the momentum and energy dependence
of inelastically scattered photons in complex materials. RIXS rapidly
became the forefront of experimental photon science
~\cite{AVD+11,GHE+24}. It combines spectroscopy and inelastic
scattering to probe the electronic structure of materials. This method
is an element- and orbital- selective X-ray spectroscopy technique,
based on a two-step, two-photon resonant process. It combines X-ray
emission spectroscopy (XES) with X-ray absorption spectroscopy (XAS)
by measuring the coherent X-ray emission at an incident X-ray photon
energy within the near edge X-ray absorption spectrum. In the first
step (X-ray absorption), an electron of the absorbing atom is
resonantly excited from a core level to an empty state. The resulting
state, called the intermediate state, carries a core hole with a very
small lifetime. In the second step (X-ray emission), the system
radiatively decays into a final state in which the core hole is filled
by another electron accompanied by photon-out emission. The
polarization of the incoming and outgoing light and the resonant
energy are involved in the RIXS process, making RIXS a simultaneous
spectroscopy and scattering technique. RIXS has a number of unique
features in comparison with other spectroscopic techniques. It covers
a large scattering phase space and requires only small sample
volumes. Because RIXS is a photon-in, photon-out technique, it is truly bulk sensitive, 
with negligible contributions from surface effects. Additionally, RIXS is polarization 
dependent, and both element- and orbital-specific~\cite{AVD+11}. A detailed comparison 
with other spectroscopic techniques can be found in a recent review~\cite{GHE+24}. 
Spectral broadening due to the short core-hole lifetime can be minimized, to produce RIXS 
spectra with high energy and momentum resolution. This enables direct probing of quasiparticles 
and their properties, such as phonons, plasmons, single magnons, and orbitons, their mutual 
interplay, as well as other many-body excitations in strongly correlated systems, including cuprates, 
nickelates, osmates, ruthenates, and iridates, which exhibit complex and exotic low-energy physics 
manifested by fine structure of excitations in energy-momentum space.

Significant progress in RIXS experiments has been achieved over the past decade. However, most 
theoretical modeling of RIXS spectra has relied on the atomic multiplet approach with empirical 
parameter adjustments. First-principles calculations of RIXS spectra remain exceedingly rare. 
Thus, the most essential point of our paper is to introduce an \emph{ab initio} theoretical 
approach for modeling RIXS spectra and to demonstrate its application to the highly nontrivial 
and complex material Ca$_2$RuO$_4$.

Extensive experimental RIXS investigations of Ca$_2$RuO$_4$ have been reported, with the key findings 
summarized below. We begin with the work of Gretarsson \textit{et al.} \cite{GSK+19}, who measured 
and characterized the RIXS spectra at the Ru $L_3$ edge. They have uncovered a series of sharp
electronic excitations at low energy in Ca$_2$RuO$_4$ at $\sim$0.05,
0.32, 0.75, and 1.0 eV.  They found strong photon polarization
dependence of the RIXS intensity at 0.32 eV with varying the incident
angle $\theta$ from 9$^{\circ}$ to 77$^{\circ}$. The authors have carried
out ionic model calculations that quantify the energy levels of Ru
$d^4$ multiplets and corresponding RIXS intensities. The ionic model
Hamiltonian which includes the intraionic Coulomb interactions, SOC,
and CEF was used. The lowest peak was interpreted as magnetic excitations in consistency with neutron and Raman scattering
measurements \cite{JKP+17,SCK+17}. The authors interpreted the excitation
at 0.32 eV as a SOC-driven $J$ = 0 $\rightarrow$ 2 transition. The peaks
at 0.75 and 1.0 eV were found to be due to Hund's-rule driven $S$ = 1
$\rightarrow$ 0 spin-state transitions, split by the tetragonal
CEF. The multiplets at higher energies (2-4 eV)  correspond to excitations from the {\tg} 
ground-state manifold of the Ru ions into the {\eg} CEF levels. Kim {\it et al.}~\cite{KKK22} present
a similar classification of the Ru $L_3$ RIXS spectrum by developing a
comprehensive theory description of the RIXS spectrum using the general
RIXS operators introduced in Ref.~\cite{KiKh17}. Analytic solutions were provided for the collective 
magnetic excitations and for their RIXS transitions in terms of the pseudospin
operators. The authors found that the $J$ = 2 transitions at 0.32 eV show
huge spectral weights and exhibit prominent polarization dependence in
agreement with the experimental observation~\cite{GSK+19}. Bertinshaw
{\it et al.}~\cite{BKS+21} present the RIXS Ru $L_3$ spectrum in
Ca$_2$RuO$_4$ in a wide energy interval up to 6.2 eV and compare it with the
corresponding spectrum in the structurally related Mott insulator
Ca$_3$Ru$_2$O$_7$.

Yamamoto {\it et al.}~\cite{YOS22} investigated the low-energy
excitation ($\le$1.5 eV) RIXS spectra in Ca$_2$RuO$_4$ using the
three-orbital Hubbard model obtained by the band-structure calculation
and applying random-phase approximation (RPA). The energy band
structure of Ca$_2$RuO$_4$ calculated by the mean-field
approximation with $U_{\rm{eff}}$ = 1.53 eV. By the fast-collision
approximation, they calculated the Ru $L_3$ RIXS spectra from the
dynamical susceptibilities. The authors show that the dispersion of the
transverse mode is clearly observed in the calculated RIXS spectra and
that the polarization dependence of the incident X-rays enables one to
distinguish between the excitations of the in-plane and out-of-plane
transverse modes.

In this study, we present a comprehensive \emph{ab initio} DFT analysis of the electronic structure and RIXS spectra of Ca$_2$RuO$_4$ over an extended energy range. We calculated the band structure and modeled the RIXS spectra using the fully relativistic, spin‐polarized Dirac linear muffin‐tin orbital (LMTO) method. To assess the impact of electron correlation, we employed the generalized gradient approximation (GGA) together with the GGA+$U$ approach. Overall, our work provides key insights into the influence of transition-metal 4$d$–O 2$p$ hybridization and properties of the band-structure on RIXS spectra not only of Ca$_2$RuO$_4$ but of 4$d$ oxides in general.

The paper is organized as follows. Section II describes the crystal structure of Ca$_2$RuO$_4$ and the computational methodology. Section III presents the results of the electronic structure of Ca$_2$RuO$_4$. Section IV discusses the theoretical investigations of XES, XAS, and XMCD spectra. Section V presents the \emph{ab initio} RIXS spectra at the Ru $L_3$ and O $K$ edges, and compares them with experimental data. Finally, Section VI summarizes our conclusions.

\section{Details of Theoretical Methodology}
\label{sec:details}

\subsection{Modeling of RIXS spectra} 

RIXS refers to the process where the
material first absorbs a photon. The system then is excited to a
short-lived intermediate state, from which it relaxes radiatively. In an experiment, one studies the X-rays emitted in this decay process. In the direct RIXS process~\cite{AVD+11} an incoming photon
with energy $\hbar \omega_{\mathbf{k}}$, momentum $\hbar \mathbf{k}$
and polarization $\bm{\epsilon}$ excites the solid from a ground state
$|{\rm g}\rangle$ with energy $E_{\rm g}$ to the intermediate state
$|{\rm I}\rangle$ with energy $E_{\rm I}$. During relaxation the
outcoming photon with energy $\hbar \omega_{\mathbf{k}'}$, momentum
$\hbar \mathbf{k}'$ and polarization $\bm{\epsilon}'$ is emitted, and
the solid is in the state $|{\rm f}\rangle$ with energy $E_{\rm f}$. A
valence electron is excited from state $\mathbf{k}$ to states
$\mathbf{k}'$ with energy $\hbar \omega = \hbar \omega_{\mathbf{k}} -
\hbar \omega_{\mathbf{k}'}$ and momentum transfer $\hbar \mathbf{q}$ =
$\hbar \mathbf{k} - \hbar \mathbf{k}'$.
The RIXS intensity can in general be presented in terms of a
scattering amplitude as~\cite{AVD+11}

\begin{eqnarray}
I(\omega, \mathbf{k}, \mathbf{k}', \bm{\epsilon}, \bm{\epsilon}')
&=&\sum_f \left| T_{fg}(\mathbf{k}, \mathbf{k}',
\bm{\epsilon}, \bm{\epsilon}', \omega_{\mathbf{k}}) \right|^2 \nonumber \\
&&\times \delta(E_f+\hbar \omega_{\mathbf{k}'}-E_g-\hbar \omega_{\mathbf{k}})\, ,
\label{I}
\end{eqnarray}
where the delta function enforces energy conservation and the
amplitude $T_{fg}(\mathbf{k}, \mathbf{k}', \bm{\epsilon},
\bm{\epsilon}', \omega_{\mathbf{k}})$ reflects which excitations are
probed and how, for instance, the spectral weights of final state
excitations depend on the polarization vectors $\bm{\epsilon}$ and
$\bm{\epsilon'}$ of the incoming and outgoing x-rays,
respectively.

Our implementation of the code for the calculation of the RIXS
intensity uses Dirac four-component basis functions~\cite{NKA+83} in
the perturbative approach~\cite{ASG97}. RIXS is a second-order
process, and its intensity is given by

\begin{eqnarray}
I(\omega, \mathbf{k}, \mathbf{k}', \bm{\epsilon}, \bm{\epsilon}')
&\propto&\sum_{\rm f}\left| \sum_{\rm I}{\langle{\rm
    f}|\hat{H}'_{\mathbf{k}'\bm{\epsilon}'}|{\rm I}\rangle \langle{\rm
    I}|\hat{H}'_{\mathbf{k}\bm{\epsilon}}|{\rm g}\rangle\over
  E_{\rm g}-E_{\rm I}} \right|^2 \nonumber \\ && \times
\delta(E_{\rm f}-E_{\rm g}-\hbar\omega),
\label{I1}
\end{eqnarray}
where the delta function enforces energy conservation, and the photon
absorption operator in the dipole approximation is given by the
lattice sum
$\hat{H}'_{\mathbf{k}\bm{\epsilon}}=
\sum_\mathbf{R}\hat{\bm{\alpha}}\bm{\epsilon} \exp(-{\rm
  i}\mathbf{k}\mathbf{R})$, where $\hat{\bm{\alpha}}$ are Dirac
matrices. Both $|{\rm g}\rangle$ and $|{\rm f}\rangle$ states are
dispersive so the sum over final states is calculated using the linear
tetrahedron method~\cite{LeTa72}. The matrix elements of the RIXS
process in the frame of the fully relativistic Dirac LMTO method were
presented in our previous publication~\cite{AKB22a}.

\subsection{Crystal structure} 

Ca$_2$RuO$_4$ is reported to undergo a series of phase transitions
upon cooling; a metal-to-insulator transition at 357~K
~\cite{BAN+98,ACD+99,FBA+01,SFA+05}, an orbital ordering at 260~K
~\cite{ZSN+05,PGF+18}, and an AFM ordering at 110~K
~\cite{BAN+98,ACD+99,ZSN+05,PGF+18}. At low temperature below 180~K
Ca$_2$RuO$_4$ is crystallized in the simple orthorhombic structure (space
group $Pbca$, No. 61) which is characterized by a rotation of the
octahedra around the $c$ axis with a tilt around an axis parallel to
an edge of the octahedron basal plane~\cite{FBA+01}.

\begin{figure}[tbp!]
\begin{center}
\includegraphics[width=0.90\columnwidth]{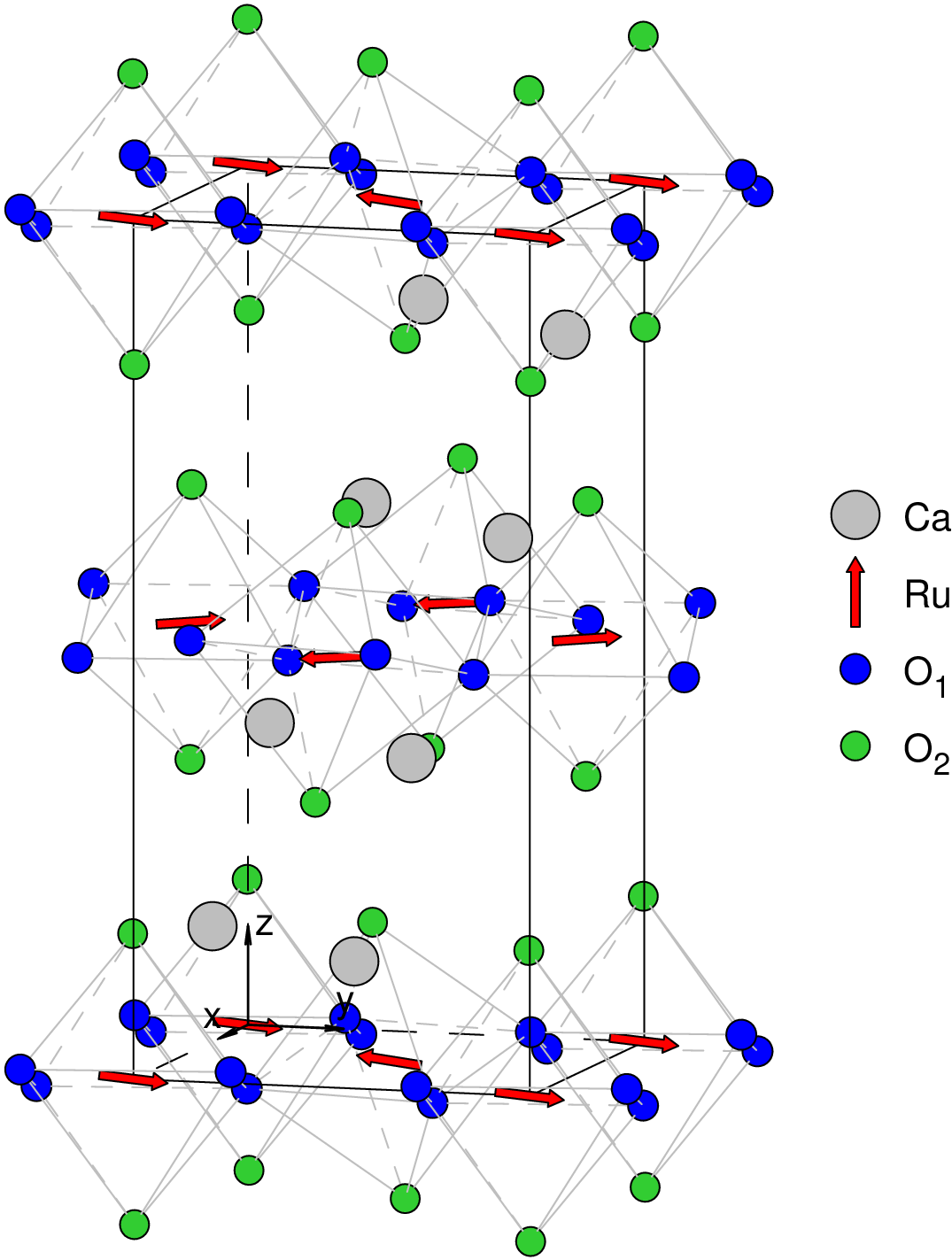}
\end{center}
\caption{\label{struc_CRO}(Color online) Schematic representation of
  the AFM structure of Ru ions (red arrows) in
  Ca$_2$RuO$_4$ (space group $pbca$, number 61). Grey spheres
  represent Ca atoms, blue and green spheres show oxygen atoms.  }
\end{figure}

The lattice parameters of Ca$_2$RuO$_4$ are given in Table
~\ref{struc_tab_CRO}. The RuO$_6$ octahedra are tilted and rotated, as
seen in Fig.~\ref{struc_CRO}. Each RuO$_6$ octahedron is flattened
since the bond distance between Ru and apical oxygen Ru-O$_2$ (=
1.9729~\AA) is shorter than the in-plane Ru-O$_1$ bond distances (=
2.0040~\AA).

\begin{table}[tbp!]
  \caption {Atomic positions in Ca$_2$RuO$_4$ ($Pbca$). The lattice parameters for Ca$_2$RuO$_4$ are equal to $a$ = 5.3945~\AA, $b$ = 5.5999~\AA\, $c$ = 11.7653~\AA\,~\cite{FBA+01}.  }
\label{struc_tab_CRO}
\begin{center}
\begin{tabular}{|c|c|c|c|c|}
\hline
 Atom & WC  & $x$      & $y$    & $z$  \\
\hline    
 Ca    & 8$c$  & 0.0042  & 0.0559 & 0.3524   \\
 Ru    & 4$a$  & 0       & 0      &  0    \\
 O$_1$ & 8$c$  & 0.1961  & 0.3018 & 0.0264   \\
 O$_2$ & 8$c$  & -0.0673 &-0.0218 & 0.1645   \\
\hline    
\end{tabular}
\end{center}
\end{table}

\subsection{Details of Computations}

The details of the computational method are described in our previous
papers~\cite{AJY+06,AHY+07b,AYJ10,AKB22a} and here we only mention
several aspects. The band structure calculations were performed using
the fully relativistic linear muffin-tin orbital (LMTO) method
~\cite{And75,book:AHY04}. This implementation of the LMTO method uses
four-component basis functions constructed by solving the Dirac
equation inside an atomic sphere~\cite{NKA+83}. The
exchange-correlation functional of the generalized gradient
approximation (GGA)-type was used in the version of Perdew, Burke and
Ernzerhof~\cite{PBE96}. The Brillouin zone (BZ) integration was performed using the improved tetrahedron method~\cite{BJA94}. The basis consisted of Ru and Ca $s$, $p$, $d$, and $f$; and O $s$, $p$, and $d$ LMTO's.

To take into account the electron-electron correlation effects, we used in this work the relativistic generalization of the rotationally
invariant version of the LSDA+$U$ method~\cite{YAF03} which takes into account that in the presence of SOC the occupation matrix of localized electrons becomes non-diagonal in spin
indexes. Hubbard $U$ was considered as an external parameter and
varied from 0.7~eV to 3.7~eV.  We used in our calculations the value
of exchange Hund's coupling $J_H$=0.7 eV obtained from constrained LSDA
calculations~\cite{DBZ+84,PEE98}. Thus, the parameter
$U_{\rm{eff}}=U-J_H$, which roughly determines the splitting between
the lower and upper Hubbard bands, varied between 0~eV and 3.0~eV. We
adjusted the value of $U$ to achieve the best agreement with the
experiment.

In the RIXS process, an electron is promoted from a core level to an intermediate state, 
leaving behind a core hole. As a result, the electronic structure of that intermediate state 
differs from that of the ground state. In order to reproduce the experimental spectrum the self-consistent calculations should be carried out including a core hole.  Usually, the
core-hole effect has no impact on the shape of XAS spectra at the
$L_{2,3}$ edges of the 5$d$ systems and just a minor impact on the XMCD
spectra at these edges~\cite{book:AHY04}. However, the core hole has a strong impact on the RIXS spectra in transition metal compounds
~\cite{AKB22a,AKB22b}, therefore, we take it into account.

The XAS, XMCD, and RIXS spectra were calculated taking into account
the exchange splitting of core levels. The finite lifetime of a core
hole was accounted for by folding the spectra with a Lorentzian. The
widths of core levels for Ru and O were taken from Ref.~\cite{CaPa01}. The finite experimental resolution of the
spectrometer was accounted for by a Gaussian of 0.6~eV (the $s$
coefficient of the Gaussian function).

Note that in our electronic structure calculations, we rely on
experimentally measured atomic positions and lattice parameters
~\cite{FBA+01} (see Table~\ref{struc_tab_CRO}) because they are well
established for these materials and are probably still more accurate
than those obtained from DFT.

\section{Electronic structure}
\label{sec:bands}

In order to gain insight about the ground state magnetic
configurations of Ca$_2$RuO$_4$ we compared the calculated total
energies for different spin ordering in Ca$_2$RuO$_4$, namely,
nonmagnetic (NM), ferromagnetic (FM), and AFM. For AFM phases we
consider three types of ordering along the $c$ (AFM$_{001}$), $a$
(AFM$_{100}$), and $b$ (AFM$_{010}$) axes. We also consider 
possible noncollinear (NC) magnetic structures. Our GGA+SO band
structure calculations show that the canted NC configuration shown 
in Fig. \ref{struc_CRO} with Ru spins ordered antiferromagnetically
almost along the (010) direction (AFM$^{\rm{NC}}_{010}$) possesses 
the lowest total energy in comparison with the NM,
FM, AFM ordering along the $a$, $b$, and $c$ directions (Table
\ref{Etot_CRO}). The NC polar angles equal to
$\theta_{\rm{Ru}}$=95.4$^{\circ}$ and
$\phi_{\rm{Ru}}$=$\pm$ 95.8$^{\circ}$. Similar results are obtained also
in the GGA+SO+$U$ approximation, the results are found to be robust
against changes in the parameter $U_{\rm{eff}}$. 
This is consistent with the conclusion made by Porter \textit{et al.}~\cite{PGF+18} regarding AFM canting in Ca$_2$RuO$_4$, based on detailed resonant elastic x-ray scattering (REXS) measurements at the Ru $L_{2,3}$ edges. Their results indicate that the magnetic moment is not strictly confined to the $b$ axis, as previously suggested, but instead has a component along the $c$ axis amounting to approximately one tenth of the total moment~\cite{PGF+18}.

\begin{table}[tbp!]
  \caption{\label{Etot_CRO} The total energy $E_{total}$ per formula
    unit (in meV) calculated in the GGA approximation for the
    nonmagnetic configuration, and in the GGA+SO approximation for the nonmagnetic configuration,
    ferromagnetic ordering along the (001) axis (FM$_{001}$),
    antiferromagnetic (AFM) ordering along the (001) axis
    (AFM$_{001}$), AFM ordering along the (100) axis (AFM$_{100}$),
    AFM ordering along the (010) axis (AFM$_{010}$), and canted
    noncollinear AFM ordering almost along the (010) axis (AFM$^{\rm{NC}}_{010}$)
    defined relative to AFM$^{\rm{NC}}_{010}$. }
\begin{center}
\begin{tabular}{ccccccccccc}
\hline
GGA &  GGA+SO & FM$_{001}$  & AFM$_{001}$ & AFM$_{100}$ & AFM$_{010}$ & AFM$^{\rm{NC}}_{010}$ \\
\hline
 12.647   & 3.383 & 7.652 & 1.120 & 0.329 & 0.098 & 0.0  \\
\hline
\end{tabular}
\end{center}
\end{table}

\begin{figure}[tbp!]
\begin{center}
\includegraphics[width=0.99\columnwidth]{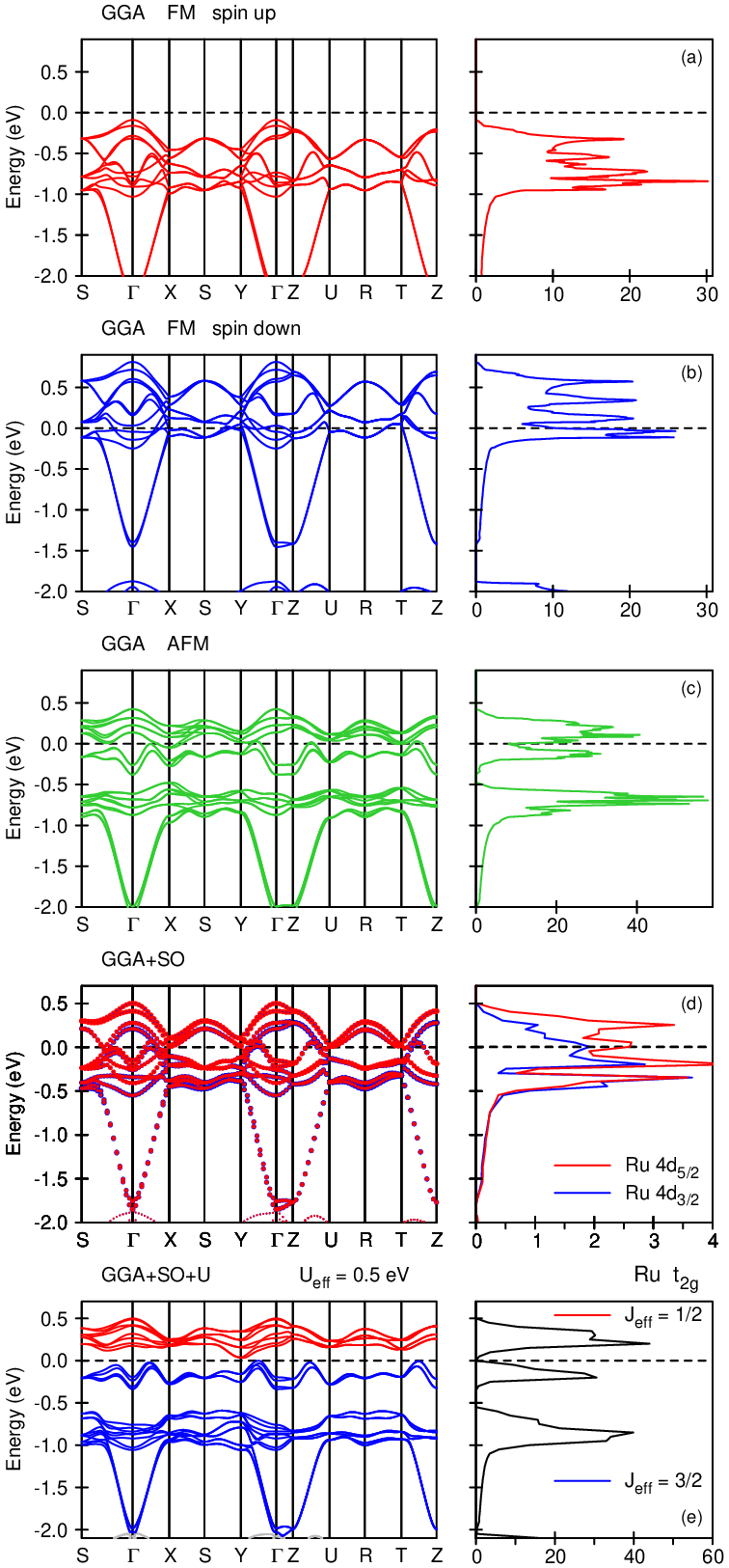}
\end{center}
\caption{\label{BND_Jeff_CRO}(Color online) The {\tg} energy band
  structure of Ca$_2$RuO$_4$ calculated in the GGA approach without
  SOC for the FM ordering for the spin up (a) and spin down (b)
  states. The panel (c) shows the energy bands for the AFM ordering in
  the GGA approach. The panel (d) presents the fully relativistic
  Dirac GGA+SO approximation. The bands crossing the Fermi level which
  have almost pure $d_{5/2}$ character (open red circles) are formed
  by {\tg} states with $J_{\rm{eff}}$ = 1/2. The lower panel (e)
  presents the {\tg} energy bands calculated in the GGA+SO+$U$
  approximation with $U_{\rm{eff}}$ = 0.5 eV for the canted
  noncollinear AFM ordering AFM$^{\rm{NC}}_{010}$. }
\end{figure}

Figure \ref{BND_Jeff_CRO} establishes a picture of the SOC-driven Mott
transition in Ca$_2$RuO$_4$. In the absence of SOC, the partially
filled bands of predominantly {\tg} orbital character would lead to a
metallic ground state [Fig. \ref{BND_Jeff_CRO}(a,b,c)]. The GGA
approach for the FM ordering produces a half-metallic solution with
the energy gap for spin up states. The Fermi energy is located near
a local minimum in the DOS for the AFM solution
[Fig. \ref{BND_Jeff_CRO}(c)].

The fully relativistic GGA+SO bands are presented in
Fig. \ref{BND_Jeff_CRO}(d) by circles proportional in size to their
orbital character projected onto the basis set of Ru $d_{3/2}$ (the
relativistic quantum number $\kappa$ = 2, the blue curve) and
$d_{5/2}$ ($\kappa$ = $-$3, the red curve) states. In Ca$_2$RuO$_4$
each Ru$^{4+}$ ion surrounded by six O$^{2-}$ ions has four valent
4$d$ electrons. The octahedral crystal field largely splits the Ru
$t_{2g}$ and $e_g$ manifolds, so that all four electrons occupy the
$t_{2g}$ manifold. As a result of strong SOC, the six $t_{2g}$
orbitals are further separated into two manifolds with $J_{\rm{eff}}$
= 3/2 and $J_{\rm{eff}}$ = 1/2. The $J_{\rm{eff}}$ = 3/2 states are
fully filled and $J_{\rm{eff}}$ = 1/2 states are empty, which is
consistent with our expectations. The functions of the $J_{\rm{eff}}$
= 3/2 quartet are dominated by $d_{3/2}$ states with some minor weight
of $d_{5/2}$ ones, which is determined by the relative strengths of
SOC and crystal-field splitting. The $J_{\rm{eff}}$ = 1/2 functions,
on the other hand, are given by linear combinations of $d_{5/2}$
states only [Fig. \ref{BND_Jeff_CRO}(d)]. This allows one to identify
bands with pure $d_{5/2}$ character as originating from $J_{\rm{eff}}$
= 1/2 states.

The lower panel (e) of Fig. \ref{BND_Jeff_CRO} presents the {\tg}
energy bands calculated in the GGA+SO+$U$ approximation with
$U_{\rm{eff}}$ = 0.5 eV for the canted noncollinear AFM ordering
(AFM$^{\rm{NC}}_{010}$) with the energy gap between fully occupied
$J_{\rm{eff}}$ = 3/2 and empty $J_{\rm{eff}}$ = 1/2 states.

It is useful to compare Ca$_2$RuO$_4$ with Sr$_2$IrO$_4$, as the latter is less correlated due to its 5$d$ electrons, compared to the 4$d$ electrons in Ca$_2$RuO$_4$. Because of this, Sr$_2$IrO$_4$ would be expected to show a weaker tendency toward magnetism. On the other hand, Sr$_2$IrO$_4$
has only a single hole in the {\tg} manifold, as opposed to two in
Ca$_2$RuO$_4$ (5$d^5$ versus 4$d^4$ electrons, respectively), and
stronger spin-orbit interaction. In Sr$_2$IrO$_4$ for the nonmagnetic
solution the energy gap does not open up for any, even very large,
Hubbard parameter $U$ \cite{AKB24a}. In the case of Ca$_2$RuO$_4$ for
the nonmagnetic solution the insulating gap opens up at the critical
value for the $U^c_{\rm{eff}}$ = 2.05 and 1.4 eV in the GGA (without
SOC) and GGA+SO (with SOC) approximations,
respectively. For the FM ordering the gap opens up at $U^c_{\rm{eff}}$
= 1.2 eV for the GGA+SO approach. For the collinear AFM ordering the
gap opens up at smaller $U^c_{\rm{eff}}$ = 1.0 eV. For the ground
state noncollinear canted AFM$^{\rm{NC}}_{010}$ ordering the gap opens
up for even smaller $U^c_{\rm{eff}}$ = 0.35 eV. We can conclude that
magnetic ordering plays an important role in the gap formation in
Ca$_2$RuO$_4$, therefore, this oxide has a mixed Slater and Mott
character.

\begin{figure}[tbp!]
\begin{center}
\includegraphics[width=0.90\columnwidth]{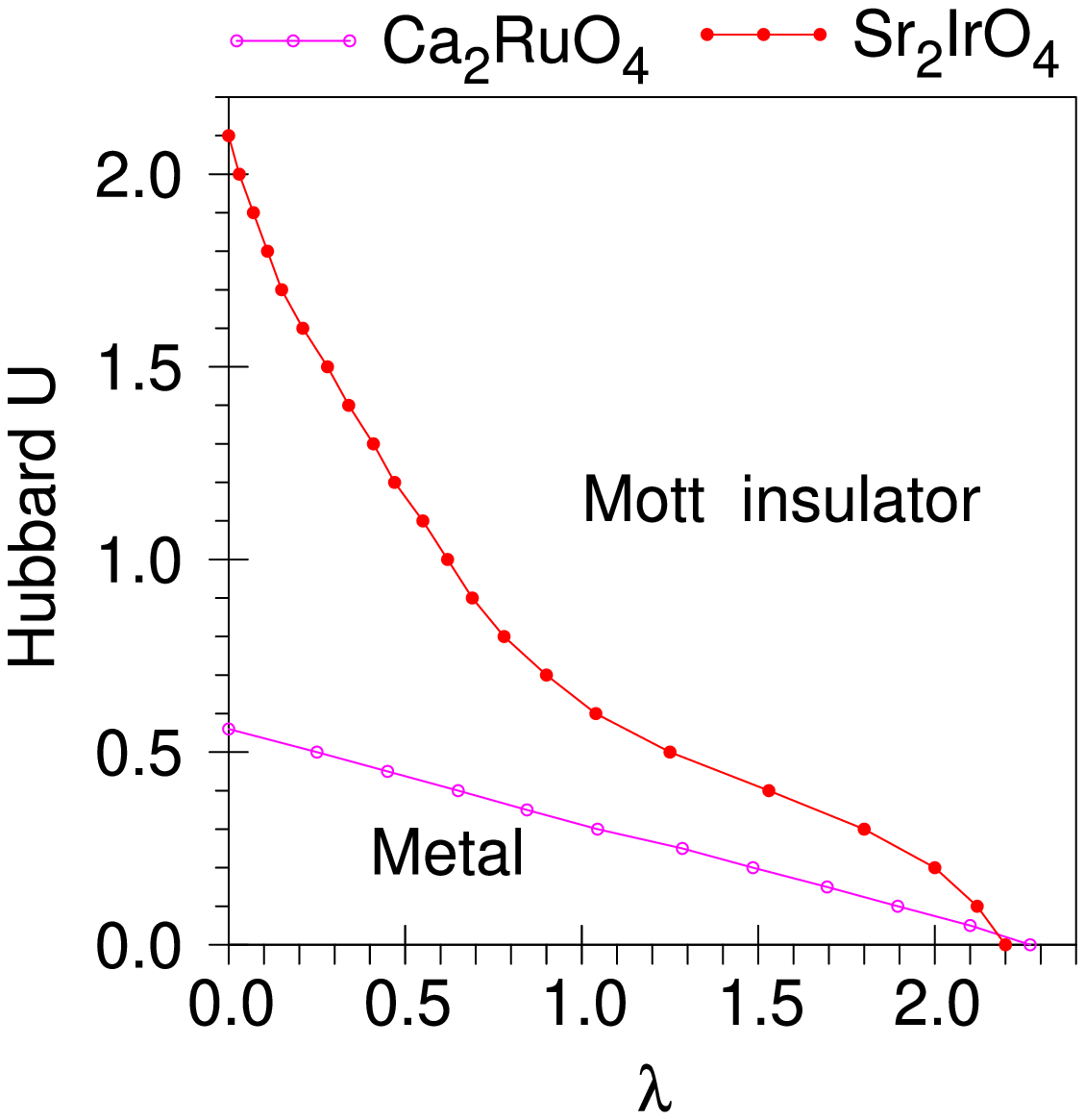}
\end{center}
\caption{\label{U_SO_CRO}(Color online) The phase diagram in
  the $U_{\rm{eff}}-$SOC plane for Ca$_2$RuO$_4$ in comparison with
  Sr$_2$IrO$_4$ \cite{AKB24a}. The solid-circled lines separates metal
  and Mott insulator states connected via a first-order phase
  transition. }
\end{figure}

Figure \ref{U_SO_CRO} presents a phase diagram in the
$U_{\rm{eff}}-$SOC plane for Ca$_2$RuO$_4$ in comparison with
Sr$_2$IrO$_4$ \cite{AKB24a}. To obtain this diagram, we tune the SOC
term for the Ru 4$d$ orbitals. A scaling factor $\lambda$ in the SOC
term of the Hamiltonian is introduced in the second variational step
\cite{KoHa77}. In this way, we can enhance the effect of SOC by taking
$\lambda$ $>$ 1 or reduce it by taking $\lambda$ $<$ 1. For $\lambda$
= 0 there is no SOC at all, while $\lambda$ = 1 refers to the
self-consistent reference value. The open-circled magenta line in
Fig. \ref{U_SO_CRO} separates metal and Mott insulator states for
Ca$_2$RuO$_4$, which are connected via a first-order phase transition,
calculated in the GGA+SO+$U$ approximation for the canted
AFM$^{\rm{NC}}_{010}$ order. For $\lambda$ = 0 the energy gap opens up for
$U^c_{\rm{eff}}$ = 0.56 eV and $U^c_{\rm{eff}}$ = 0 eV for $\lambda$ = 2.27. 
The greater the value of $U_{\rm{eff}}$, the lower the value of $\lambda$ is for the phase
transition.

In the case of Sr$_2$IrO$_4$, for $\lambda$ = 0 the energy gap opens up for
$U^c_{\rm{eff}}$ = 2.1 eV and $U^c_{\rm{eff}}$ = 0 eV for $\lambda$ = 2.2
(the solid-circled red line in Fig. \ref{U_SO_CRO}) \cite{AKB24a}. We can conclude 
that Ca$_2$RuO$_4$ is a more strong Mott dielectric than Sr$_2$IrO$_4$ because it possesses
much larger $U_{\rm{eff}}-$SOC phase space in comparison with
Sr$_2$IrO$_4$.

\begin{figure}[tbp!]
\begin{center}
\includegraphics[width=1\columnwidth]{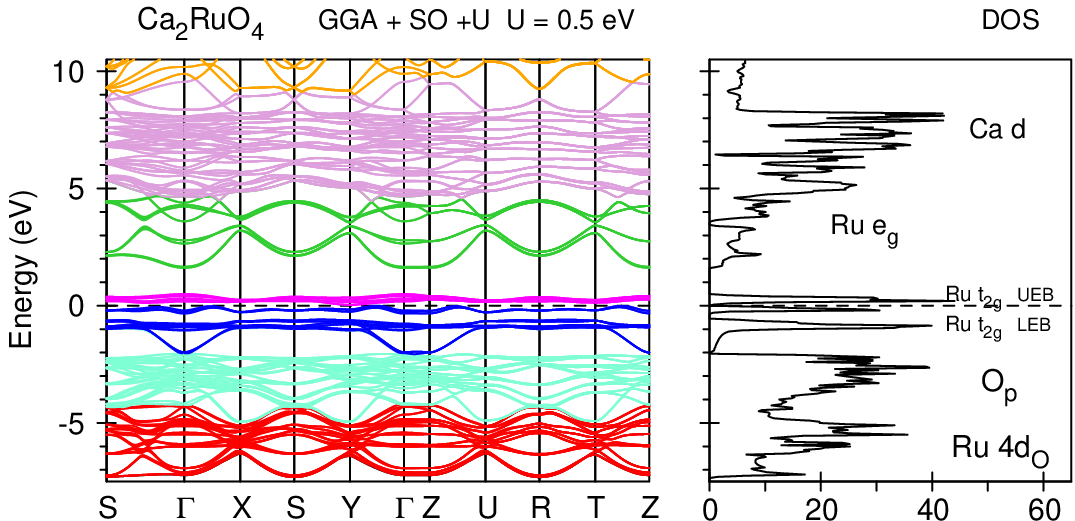}
\end{center}
\caption{\label{BND_CRO}(Color online) The {\it ab initio} energy
  band structure of Ca$_2$RuO$_4$ calculated in the GGA+SO+$U$ with
  $U_{\rm{eff}}$ = 0.5 eV approximation. }
\end{figure}

\begin{figure}[tbp!]
\begin{center}
\includegraphics[angle=-90, width=0.9\columnwidth]{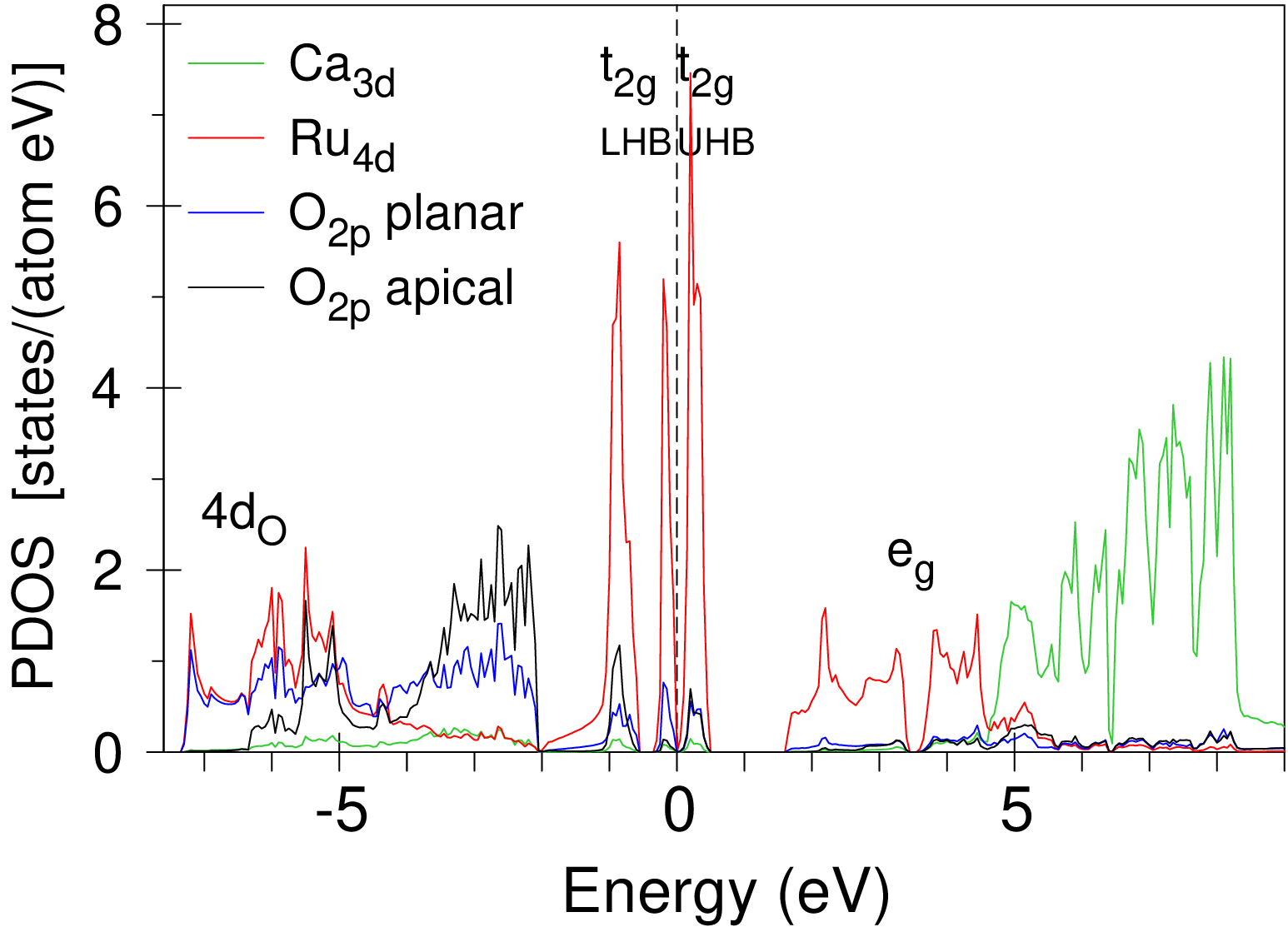}
\end{center}
\caption{\label{PDOS_CRO}(Color online) The partial density of states
  DOS [in states/(atom eV)] in Ca$_2$RuO$_4$ calculated in the
  GGA+SO+$U$ approximation ($U_{\rm{eff}}$ = 0.5 eV). }
\end{figure}

Figures \ref{BND_CRO} and \ref{PDOS_CRO} present the {\it ab initio}
energy band structure and partial DOSs of Ca$_2$RuO$_4$ for the
noncollinear canted AFM$^{\rm{NC}}_{010}$ ordering, calculated in the
fully relativistic Dirac approximation with taking into account
Coulomb correlations in the GGA+SO+$U$ approximation for 
$U_{\rm{eff}}$ = 0.5 eV. This value of Hubbard $U$ produces the best
agreement between the theoretically calculated and experimentally
measured RIXS spectra at the Ru $L_3$ and oxygen $K$ edges (see 
Section V).

Four electrons occupy the {\tg}-type low energy band (LEB) manifold
which consists of two separate peaks situated in the energy intervals
from $-$2.0 to $-$0.55 eV and from $-$0.34 eV to $E_F$. The empty
{\tg} states [the upper energy band (UEB)] have one DOS peak and
occupy the energy range from 0.04 eV to 0.5 eV (see
Fig. \ref{PDOS_CRO}). The {\eg}-type states of Ru are distributed far
above the Fermi level from 1.6 eV to 4.6 eV. The 3$d$ states of Ca
ions are mostly situated above the Fermi level from 4.6 to 9.1 eV. The
occupation number of 4$d$ electrons in the Ru atomic sphere in
Ca$_2$RuO$_4$ is equal to 5.77, which is much larger than the expected
value of four {\tg} electrons. The excessive charge is provided by the
tails of oxygen 2$p$ states. These 4$d_{\rm{O}}$ states are located at
the bottom of oxygen 2$p$ states from $-$7.3 eV to $-$4.1 eV and play
an essential role in the RIXS spectrum at the Ru $L_3$ edge (see
Section V).

The electronic structures of apical O$_2$ and in-plane O$_1$ ions
significantly differ from each other. The apical O$_2$ 2$s$ states
consist of four very narrow peaks situated at $-$17.5 and $-$16.7
eV. The in-plane O$_1$ 2$s$ states possess a relatively wider two peak
structure from $-$18.9 to $-$17.7 eV (not shown). The oxygen 2$p$
states are situated just below Ru LEB between $-$2.1 and $-$7.3 eV.
The in-plane O$_1$ 2$p$ states hybridize with Ru 4$d$ states in the energy
interval from $-$7.3 to $-$4.1 eV, while the apical O$_2$ 2$p$ states
hybridize with Ru 4$d$ states in the smaller energy interval from $-$6.4
to $-$4.1 eV. The small peaks in the close vicinity of the Fermi level
from $-$2.0 to $E_F$ and from 0.04 eV to 0.5 eV are due to the strong
hybridization between O 2$p$ and Ru {\tg} LEB and UEB, respectively.

The theoretically calculated spin $M_s$, orbital $M_l$, and total
$M_{total}$ magnetic moments using the GGA+SO+$U$ approach
($U_{\rm{eff}}$ = 0.5 eV) for the canted AFM$^{\rm{NC}}_{010}$
solution are equal to 1.2718 {\mb}, 0.2643 {\mb}, and 1.5361 {\mb},
respectively. The spin and orbital magnetic moments at the Ca site are
relatively small ($M_s$ = 0.0054 {\mb} and $M_l$ = 0.0008 {\mb}). The
magnetic moments for in-plane O$_1$ ions are equal to $M_s$ = 0.0113
{\mb}, $M_l$ = $-$0.0012 {\mb}. For the apical O$_2$ ions the magnetic
moments are relatively larger and equal to $M_s$ = 0.118 {\mb}, $M_l$
= 0.0089 {\mb}.

\section{Modelling of PES, XAS, \protect\lowercase{and} XMCD Spectra}
\label{sec:xas}

\begin{figure}[tbp!]
\begin{center}
\includegraphics[width=0.9\columnwidth]{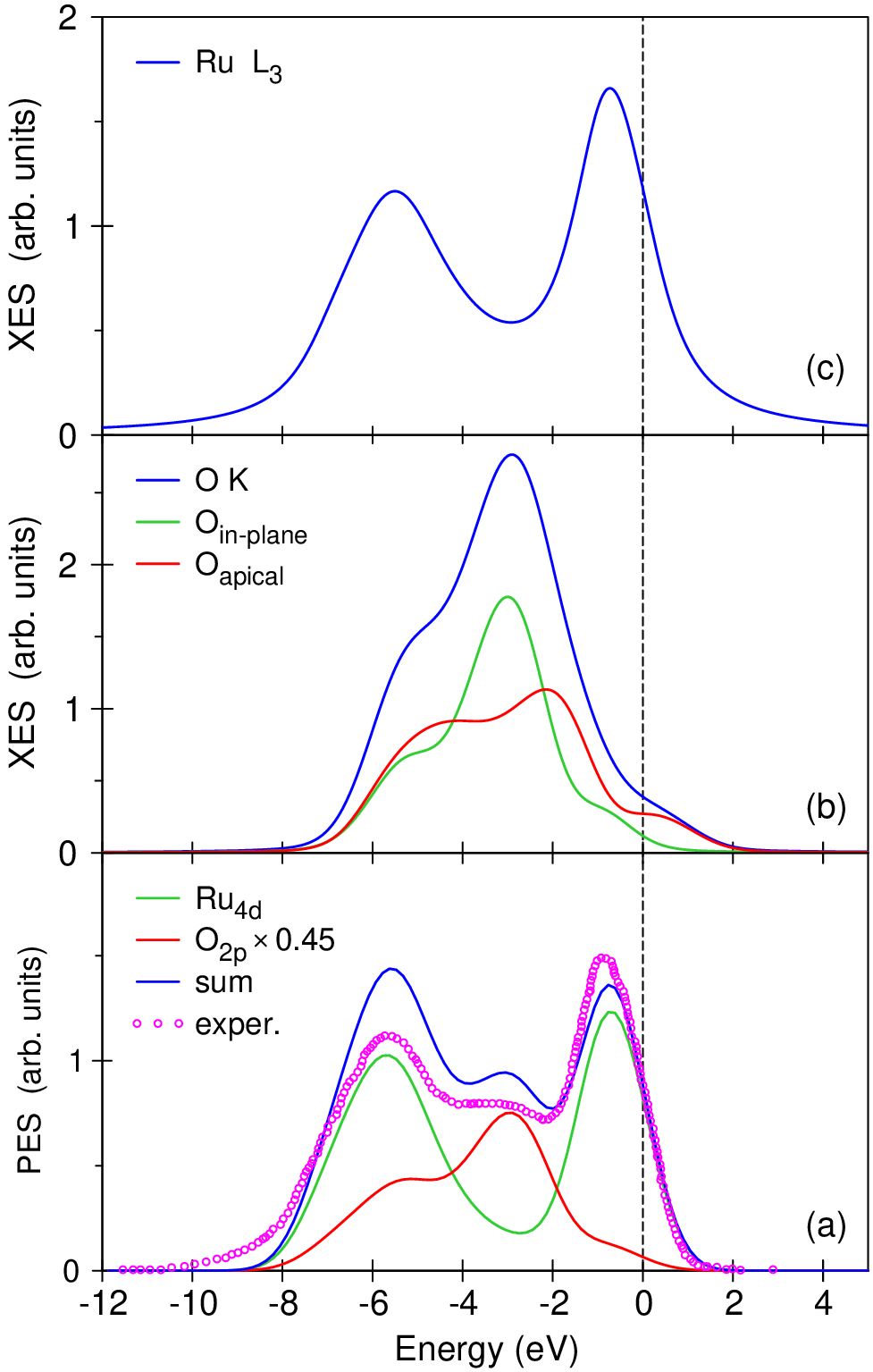}
\end{center}
\caption{\label{xps_Ru_CRO}(Color online) (a) The experimental valence-band photoemission spectrum of Ca$_2$RuO$_4$ (magenta circles) \cite{MTS+01} obtained at room temperature compared with the theoretically calculated Ru 4$d$ (full green curve) and O 2$p$ (multiply by the factor 0.45, full green curve) partial DOSs in the GGA+SO+$U$ approximation. (b) and (c) Calculated XES spectra for O $K-$ and Ru $L_3-$ edges, respectively.}
\end{figure}

We will continue with the analysis of valence band states, which can be experimentally probed by photoemission and X-ray emission spectroscopy. Figure~\ref{xps_Ru_CRO}(a) presents the experimentally obtained valence-band photoemission spectrum of Ca$_2$RuO$_4$ (magenta circles)~\cite{MTS+01}, compared with the theoretically calculated partial DOSs for Ru 4$d$ and O 2$p$ computed within the GGA+SO+$U$ approximation. The closest to the Fermi energy peak
at $\sim -$1 eV is derived mostly from the Ru {\tg} states. The peak
at $-$6 eV is due to the Ru 4$d_{\rm{O}}$ states with some
contribution from the O 2$p$ states. The fine structure at $-$3
eV is completely from the O 2$p$ states. It is known \cite{book:AHY04}
that the cross section for the 4$d$ states is larger than for the
O 2$p$ states in the valence-band PE spectra,
therefore, to achieve the best agreement with the experimental PES
spectrum one has to reduce the contribution from the O 2$p$ partial
DOS in comparison with the Ru 4$d$ states.

Figures \ref{xps_Ru_CRO}(b) and \ref{xps_Ru_CRO}(c) present the
theoretically calculated XES spectra for the oxygen $K$ and Ru $L_3$
edges, respectively. The Ru $L_3$ XES spectrum has a two peak
structure. These peaks reflect the energy position of the Ru {\tg} and
4$d_{\rm{O}}$ states similar to the corresponding peaks in the PES
spectrum. The width of the O $K$ XES spectrum is much smaller than the
Ru $L_3$ XES spectrum and the spectrum consists of a major peak at $-$3 eV with
a low energy shoulder at $\sim$ $-$5 eV. The shapes of the XES
individual spectra from the planar (the green curve) and apical (the red curve)
oxygen are quite different. Experimental measurements of the XES
at the Ru $L_3$ and oxygen $K$ edges are highly desirable.

\begin{figure}[tbp!]
\begin{center}
\includegraphics[width=0.9\columnwidth]{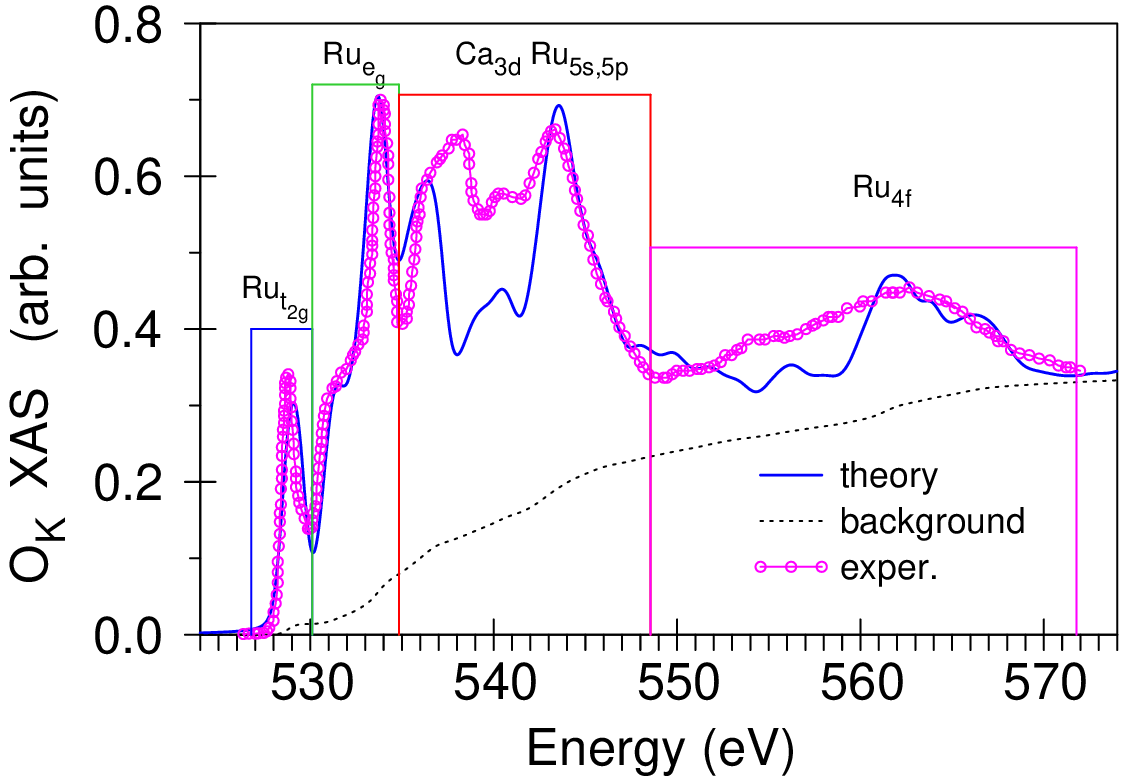}
\end{center}
\caption{\label{XAS_O_CRO}(Color online) The experimental O $K$-edge XAS spectrum of Ca$_2$RuO$_4$ taken at 90~K in TEY mode~\cite{MTS+01}, shown in comparison with the theoretical spectrum calculated within the GGA+SO+$U$ approximation using $U_{\rm{eff}} = 0.5$~eV. }
\end{figure}

X-ray absorption spectra reflect the energy distribution of empty
states in crystals. Figure \ref{XAS_O_CRO} presents the experimentally
measured oxygen $K$ XAS spectrum of Ca$_2$RuO$_4$ \cite{MTS+01} in
comparison with the theoretically calculated spectrum in the
GGA+SO+$U$ approximation in a wide energy interval. The spectrum can be
subdivided in several parts. The low energy peak at 528.8 eV is due to
dipole transitions from the 1$s$ core state into the empty oxygen 2$p$ states
derived from the hybridization with the Ru UHB {\tg} peak situated
just above the Fermi level (see Fig. \ref{PDOS_CRO}). The next peak at
534 eV with a low energy shoulder at 531 eV is due to transitions into
the oxygen 2$p$ empty states derived from the hybridization with Ru {\eg}
states. The fine structures located between 535 and 549 eV reflect the
energy distribution of the Ca$_{3d}$ and Ru$_{5s,5p}$ states. The wide
high energy peak at 549-573 eV is due to dipole transitions from the 1$s$
core state into the empty oxygen 2$p$ states derived from the
hybridization with the Ru$_{4f}$ and free electron-like
states. Although the experimental O $K$ XAS spectrum spreads over a
very wide energy interval ($\sim$45 eV), the theory quite well
describes all peculiarities of the experimental spectrum, however, it
slightly underestimates the intensity at 536-543 eV.

\begin{figure}[tbp!]
\begin{center}
\includegraphics[width=0.99\columnwidth]{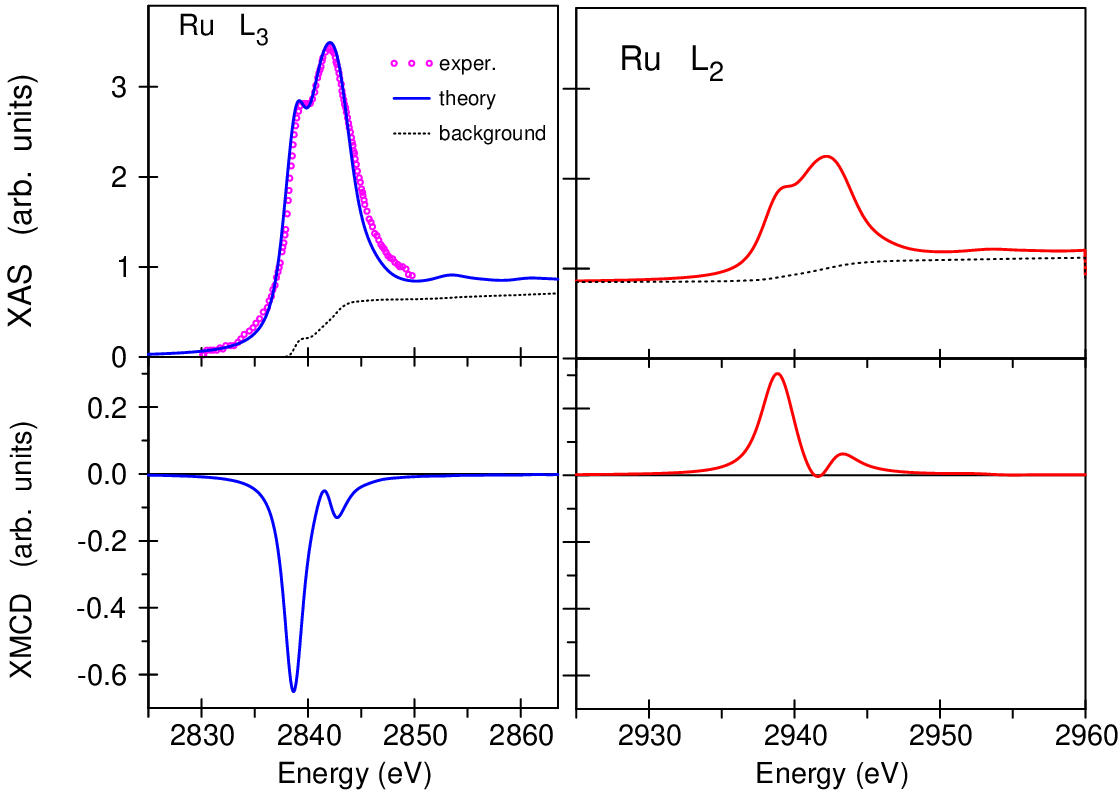}
\end{center}
\caption{\label{xmcd_Ru_CRO}(Color online) {\textbf{Upper panel.} The experimental Ru $L_3$-edge 
XAS spectrum of Ca$_2$RuO$_4$ (magenta circles) taken at room temperature~\cite{GSK+19}, shown in 
comparison with the theoretical spectrum calculated within the GGA+SO+$U$ approximation (the solid 
blue curve). The dotted black curve represents the background scattering intensity. The theoretical 
Ru $L_2$-edge XAS spectrum is also shown in red. \textbf{Lower panel.} Theoretically calculated 
Ru $L_{2,3}$ XMCD spectra.}}
\end{figure}

Figure \ref{xmcd_Ru_CRO} (the upper panel) shows the experimental
XAS spectrum at the Ru $L_3$ edge of Ca$_2$RuO$_4$
~\cite{GSK+19} in comparison with the theoretically calculated
one. The lower panel shows the theoretically calculated Ru $L_{2,3}$
XMCD spectra. Two features can be observed in the experimental Ru
$L_3$ XAS spectrum at incident energies of $E_1$ = 2838.5 eV and $E_2$
= 2841 eV, corresponding to 2$p_{3/2}$ $\rightarrow$ 4$d_{\tg}$
and 2$p_{3/2}$ $\rightarrow$ 4$d_{\eg}$ transitions, respectively
~\cite{GSK+19}. The isotropic $L_{2,3}$ XAS spectra are dominated by
the empty $e_g$ states with a smaller contribution from the empty
$t_{2g}$ orbitals at lower energy. The XMCD spectra, however, mainly
come from the $t_{2g}$ orbitals ($J_{\rm{eff}}$ = 1/2). This results
in a shift between the maxima of the XAS and XMCD spectra.

Due to the importance of SOC effects in ruthenates and iridates, it is
natural to quantify the strength of the SO interactions in these
compounds. One method of accomplishing this is provided by the XAS spectroscopy.  Van der Laan and Thole showed that the
so-called branching ratio BR = $I_{L_3}/I_{L_2}$ ($I_{L_{2,3}}$ is the
integrated intensity of the isotropic XAS at the $L_{2,3}$ edges) is
an important quantity in the study of 4$d$ and 5$d$ oxides related to
the SO interaction~\cite{LaTh88}. The BR is directly related to the
ground-state expectation value of the angular part of the spin-orbit
coupling $<{\bf L \cdot S}>$ through BR = $(2 + r)/(1 - r)$, with $r$=
$<{\bf L \cdot S}>/n_h$ and $n_h$ is the number of holes in $d$ states
~\cite{LaTh88}. As a result, XAS provides a direct probe of SO
interactions, which is complementary to other techniques such as the
magnetic susceptibility, electron paramagnetic resonance, and
M\"ossbauer spectroscopy (which probe SOC through the value of the
Lande $g$-factor). In the limit of negligible SOC effects the
statistical branching ratio BR = 2, and the $L_3$ white line is twice
the size of the $L_2$ feature~\cite{LaTh88}. A strong deviation from 2
indicates a strong coupling between the local orbital and spin
moments. Our DFT calculations produce BR = 2.46 for the GGA+SO+$U$
($U_{\rm{eff}}$ = 0.5 eV) approximation. We should mention that,
although, the BR ratio in Ca$_2$RuO$_4$ is larger than the statistical
ratio in the absence of orbital magnetization, it is still smaller in
comparison with iridates with strong SOC, such as Sr$_2$IrO$_4$, where
the measured BR is close to 4.1~\cite{HFZ+12} and the theoretically
calculated one is equal to 3.56~\cite{AKB24a}. It indicates that SOC
is less important in Ca$_2$RuO$_4$ in comparison with iridates.

\section{RIXS spectra}
\label{sec:rixs}

\subsection{Ru $L_3$ RIXS spectrum}
\label{sec:rixs_Ru}

\textcolor{black}{The RIXS spectra at the Ru $L_{2,3}$ edges arise from local excitations between the filled and empty 4$d$ states. More precisely, the incoming photon excites a 2$p_{1/2}$ core electron ($L_2$ spectrum) or a 2$p_{3/2}$ one ($L_3$ spectrum) into an empty 4$d$ state, which is subsequently followed by a de-excitation from an occupied 4$d$ state into the core level.} Because of the dipole selection rules, apart from
5$s_{1/2}$-states (which have a small contribution to RIXS due to
relatively small 2$p$ $\rightarrow$ 5$s$ matrix elements
~\cite{book:AHY04}) only 4$d_{3/2}$-states occur for $L_2$ RIXS,
whereas for $L_3$ RIXS 4$d_{5/2}$-states also contribute. Although the
2$p_{3/2}$ $\rightarrow$ 4$d_{3/2}$ radial matrix elements are only
slightly smaller than the 2$p_{3/2}$ $\rightarrow$ 4$d_{5/2}$ ones,
the angular matrix elements strongly suppress the 2$p_{3/2}$
$\rightarrow$ 4$d_{3/2}$ contribution~\cite{book:AHY04}. Therefore,
the RIXS spectrum at the Ru $L_3$ edge can be viewed as interband
transitions between 4$d_{5/2}$ states.

\begin{figure}[tbp!]
\begin{center}
\includegraphics[width=0.99\columnwidth]{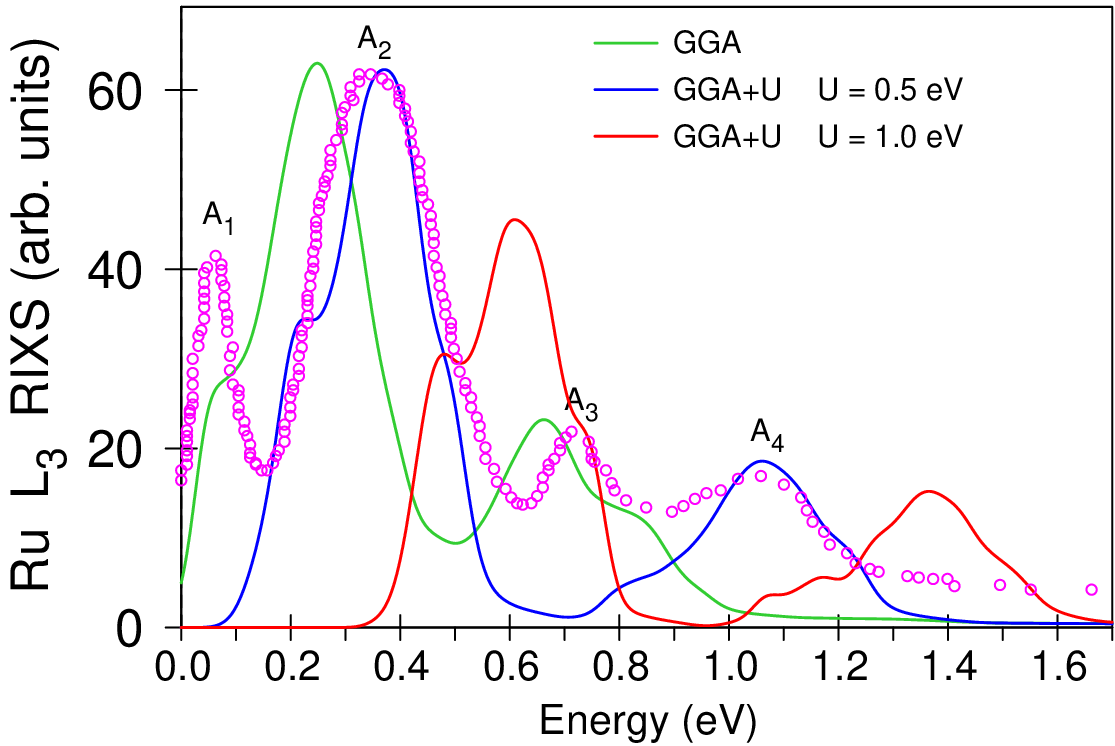}
\end{center}
\caption{\label{rixs_u_CRO}(Color online) The experimental RIXS spectrum of Ca$_2$RuO$_4$ at the Ru $L_3$ edge, obtained by Bertinshaw \textit{et al.}~\cite{BKS+21} (open magenta circles), for the $\tg \rightarrow \tg$ transitions, shown in comparison with theoretical RIXS spectra calculated using different approximations. }
\end{figure}

Ru {\tg} LEB and UEB states have two and one DOS peaks, respectively
(see Figs.~\ref{BND_CRO} and~\ref{PDOS_CRO}). The interband transitions
between these peaks produce a two peak structure in the intra-{\tg}
excitations. The energy position of these two peaks are very sensitive
to the relative position of the {\tg} LEB and UEB. Figure
\ref{rixs_u_CRO} shows the experimental RIXS spectrum obtained by
Bertinshaw {\it et al.}~\cite{BKS+21} (open magenta circles) compared
with the theoretical spectra calculated for the $\tg\rightarrow \tg$
transitions in different theoretical approaches. The GGA+SO
approximation produces the RIXS spectrum in pure agreement with the
experimental data. The best agreement was found for the GGA+SO+$U$
approximation with $U_{\rm{eff}}$ = 0.5 eV. The calculations with
larger values of $U_{\rm{eff}}$ shift the RIXS spectra towards higher
energies. The experiment produces four peaks $A_1$, $A_2$, $A_3$, and
$A_4$ at 0.05, 0.32, 0.75, and 1.0 eV (Fig.~\ref{rixs_u_CRO}) below
1.5 eV. Our calculations show a two peak structure for the
$\tg\rightarrow \tg$ transitions. There are no peaks $A_1$ and $A_3$
in the DFT calculations. The lowest peak $A_1$ at 0.05 eV was
interpreted as magnetic excitations~\cite{GSK+19} in consistency
with neutron and Raman scattering measurements~\cite{JKP+17,SCK+17}.

Gretarsson {\it et al.}~\cite{GSK+19} obtained the photon polarization
dependence of the RIXS intensity, which was modulated by angle
$\theta$ between the incoming beam and the RuO$_2$ planes of
Ca$_2$RuO$_4$. When increasing $\theta$, the polarization of the
incoming photon moves from the sample $c$ axis into the $ab$
plane. The authors used a fixed angle of 90$^{\circ}$ between incoming and
outgoing photon beams. Such investigation can provide additional
clues to the origin of the different features of the RIXS
spectrum. The authors found very strong dependence of intensity of the
$A_2$ and $A_4$ features as a function of angle $\theta$. On the other
hand, such dependence was extremely weak for the $A_3$ peak. It can
indicate that peak $A_3$ (which is absent in our DFT
calculations) possesses quite different nature in comparison with
peaks $A_2$ and $A_4$. 
\textcolor{black}{Besides, the experimental O $K$-edge RIXS spectra of Ca$_2$RuO$_4$ exhibit only two peaks below 1.5~eV~\cite{DFF+18}} It is natural to suggest that peak $A_3$
might have excitonic nature. It is interesting to note that the
reference iridate Sr$_2$IrO$_4$ also possesses four experimental peaks
for the $\tg\rightarrow \tg$ transitions~\cite{KDS+14} and only two
peaks appear in the DFT calculations~\cite{AKB24a}. The low energy
peak at 0.1 eV was considered in Ref.~\cite{KDS+14} as magnon
excitations in agreement with scanning tunneling microscope
measurements~\cite{NBA+14}. Peak at 0.5 eV, which was absent in the
DFT calculations, was attributed to an excitonic excitation in
Refs.~\cite{KDS+14,KDK+23}. The theoretical description of magnon and
exciton spectra demands a many-body approach beyond the one-particle
approximation, such as the Bethe-Salpiter equation for exciton spectra
and calculations of the magnon dispersion and the electron-magnon
interaction for magnon spectra.

\begin{figure}[tbp!]
\begin{center}
\includegraphics[width=0.9\columnwidth]{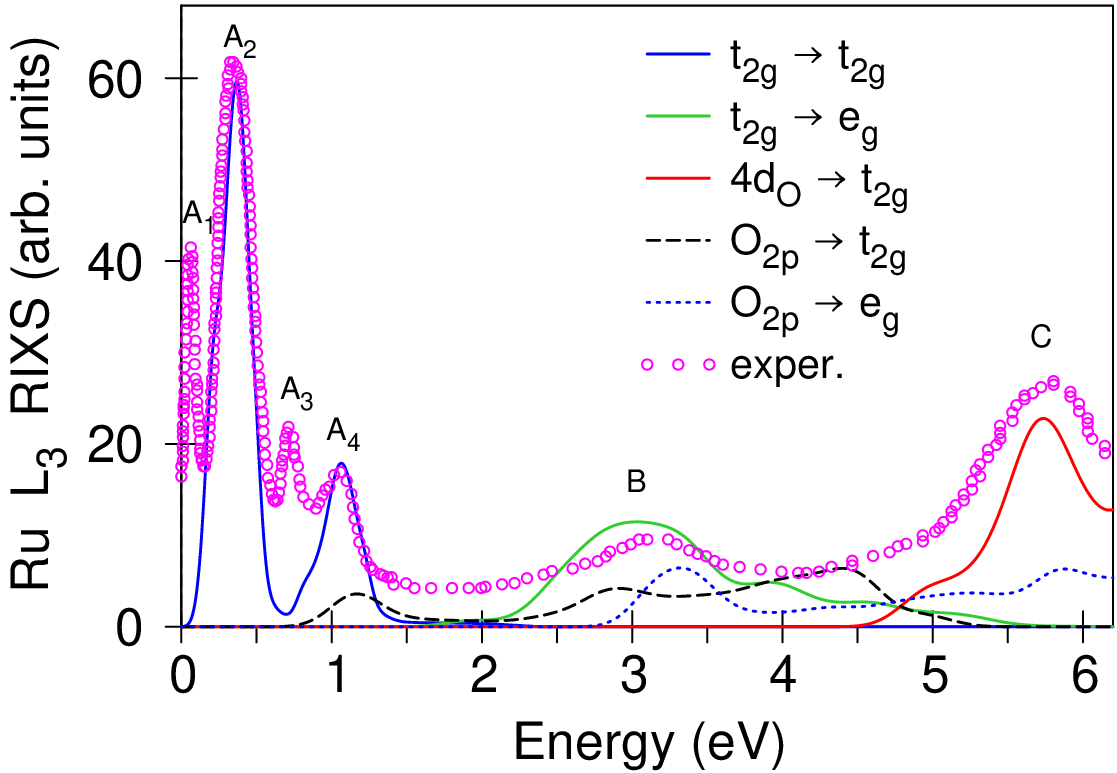}
\end{center}
\caption{\label{rixs_Ru_CRO}(Color online) The experimental RIXS
  spectrum measured by Bertinshaw {\it et al.}~\cite{BKS+21} (open
  magenta circles) at the Ru $L_3$ edge of Ca$_2$RuO$_4$ for $\theta$
  = 45$^{\circ}$ compared with the theoretically calculated partial
  contributions from different interband transitions in the GGA+SO+$U$
  ($U_{\rm{eff}}$ = 0.5 eV) approximation. }
\end{figure}

The Ru $L_3$ edge RIXS spectrum of Ca$_2$RuO$_4$ reveals several
peaks above the intra-{\tg} excitations at higher energies
(Fig.~\ref{rixs_Ru_CRO}). Peak $B$ located between 2.5 eV and 4 eV
is mostly due to $\tg \rightarrow \eg$ transitions (the green curve) with
some additional O$_{2p}$ $\rightarrow \tg$ and O$_{2p}$ $\rightarrow
\eg$ (black dashed and blue dotted curves, respectively)
transitions. Fine structure $C$ at 5 to 7 eV (the red curve) is due to
4$d_{\rm{O}}$ $\rightarrow$ {\tg} transitions.  The theoretical
calculations are in good agreement with the experimental data.

\begin{figure}[tbp!]
\begin{center}
\includegraphics[width=0.9\columnwidth]{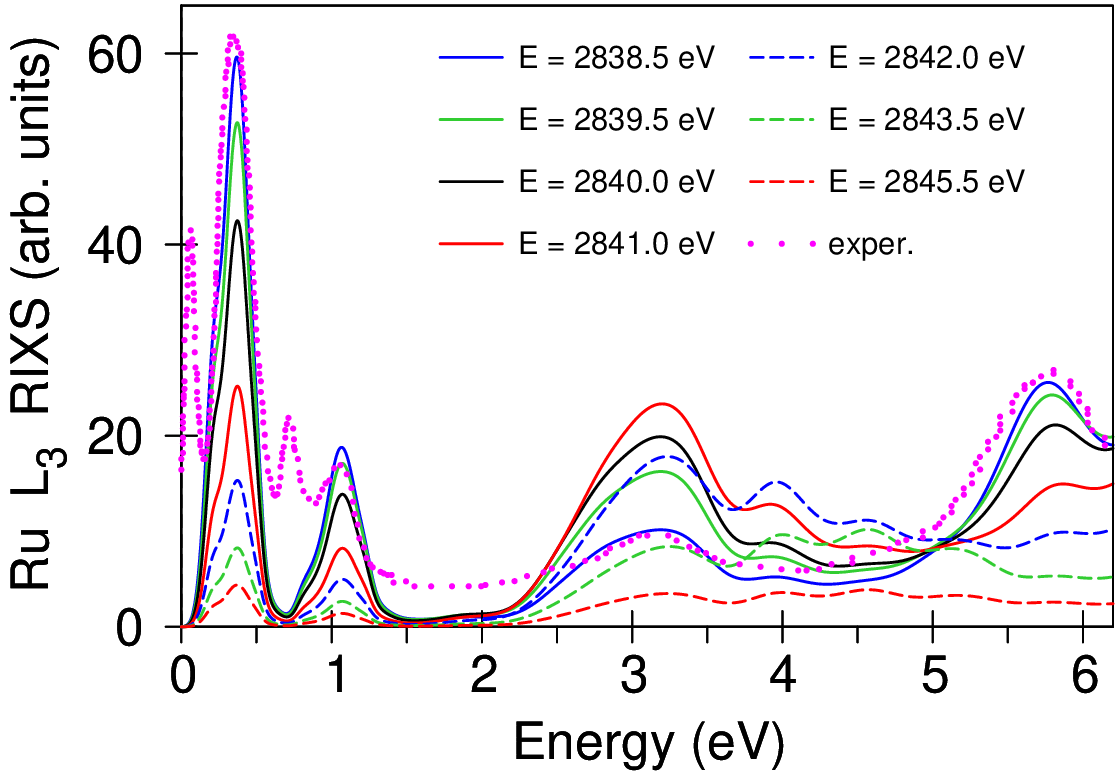}
\end{center}
\caption{\label{rixs_Ru_Ei}(Color online) The Ru $L_3$ RIXS spectra obtained as
  a function of incident photon energy $E_i$ calculated at the \textcolor{black}{Ru} $L_3$ edge in Ca$_2$RuO$_4$ with the momentum transfer vector {\bf
    Q} = (0, 0, 3.75) in reciprocal lattice units in comparison with
  the experimental RIXS spectrum measured for for $\theta$ =
  45$^{\circ}$ \cite{BKS+21}. }
\end{figure}

Figure \ref{rixs_Ru_Ei} shows the Ru $L_3$ RIXS spectra obtained as a function
of incident photon energy $E_i$ calculated in Ca$_2$RuO$_4$ for
$\theta$ = 45$^{\circ}$ with the momentum transfer vector {\bf Q} =
(0, 0, 3.75) in reciprocal lattice units in comparison with the
experimental RIXS spectrum measured for $\theta$ = 45$^{\circ}$
\cite{BKS+21}. Peak $A_2$ is monotonically decreased with increasing
the incident photon energy from $E_1$ = 2838.5 eV which corresponds to
the 4$d_{\tg}$ edge. On the other hand, the intensity of peak $B$ at
2.5-4 eV which is derived from the 2$p_{3/2}$ $\rightarrow$ 4$d_{\eg}$
transitions is increased from $E_1$ to $E_2$ = 2841 eV (which
corresponds to the 4$d_{\eg}$ edge) and then decreased with the further increase of energy.

\begin{figure}[tbp!]
\begin{center}
\includegraphics[width=0.9\columnwidth]{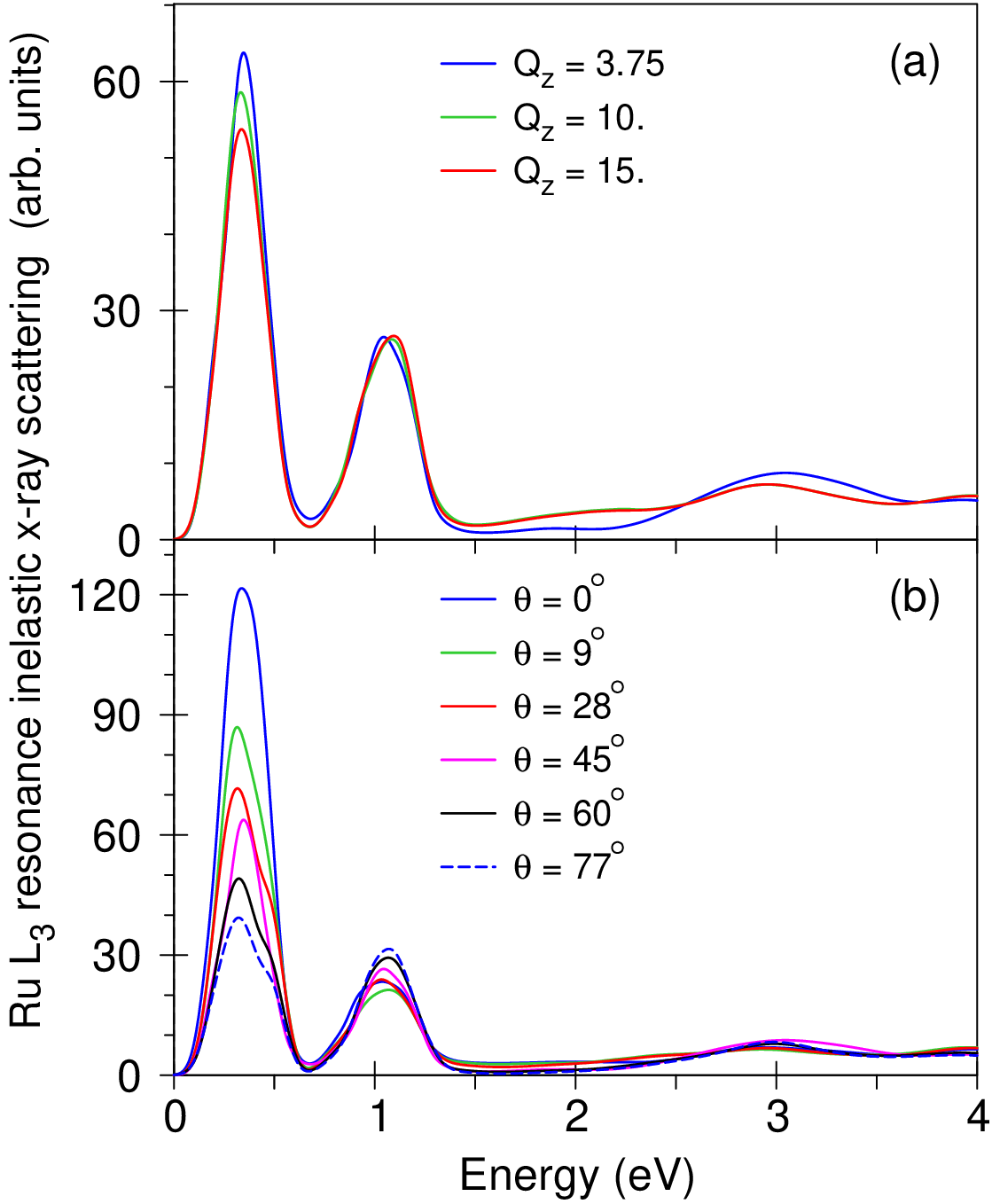}
\end{center}
\caption{\label{rixs_Ru_Qz_angle}(Color online) (a) The RIXS spectra
  at the Ru $L_3$ edge in Ca$_2$RuO$_4$ calculated as a function of
  $Q_z$ with the momentum transfer vector {\bf Q} = (0, 0, Q$_z$) in
  reciprocal lattice units for incident photon energy $\hbar
  \omega_{in}$ = 2838.5 eV ($\theta$ = 45$^{\circ}$); (b) the RIXS
  spectra at the Ru $L_3$ edge of Ca$_2$RuO$_4$ calculated as a
  function of $\theta$. }
\end{figure}

It is widely believed that $d-d$ excitations show only small momentum
transfer vector {\bf Q} dependence in $d$ transition metal compounds
\cite{LKH+12,KTD+20}. In particular, Ca$_2$RuO$_4$ has a
layered-perovskite structure, therefore, the momentum dependence along
the $c$ axis is expected to be small, as in high-$T_c$ cuprates
\cite{ITE+05}. Indeed, as we see in the upper panel of
Fig. \ref{rixs_Ru_Qz_angle}, the RIXS spectra are almost identical for
the transfer vectors {\bf Q} = (0, 0, 3.75), (0, 0, 10), and (0, 0,
20). Similar dependence was experimentally observed also in quasi-two
dimensional Sr$_2$IrO$_4$ by Ishii {\it et al.} \cite{IJY+11}.

\textcolor{black}{Figure~\ref{rixs_Ru_Qz_angle}(b) shows the RIXS spectra at the Ru $L_3$ edge for Ca$_2$RuO$_4$, calculated as a function of the angle $\theta$, and reveals their very strong polarization dependence.} The intensity of peak $A_2$ at 0.32 eV is strongly decreased with decreasing $\theta$ from 0$^{\circ}$ to 90$^{\circ}$. Very similar dependence was observed experimentally by Gretarsson {\it et
  al.}~\cite{GSK+19}.

\subsection{Oxygen $K$ RIXS spectrum}
\label{sec:rixs_O_K}

\begin{figure}[tbp!]
\begin{center}
\includegraphics[width=0.9\columnwidth]{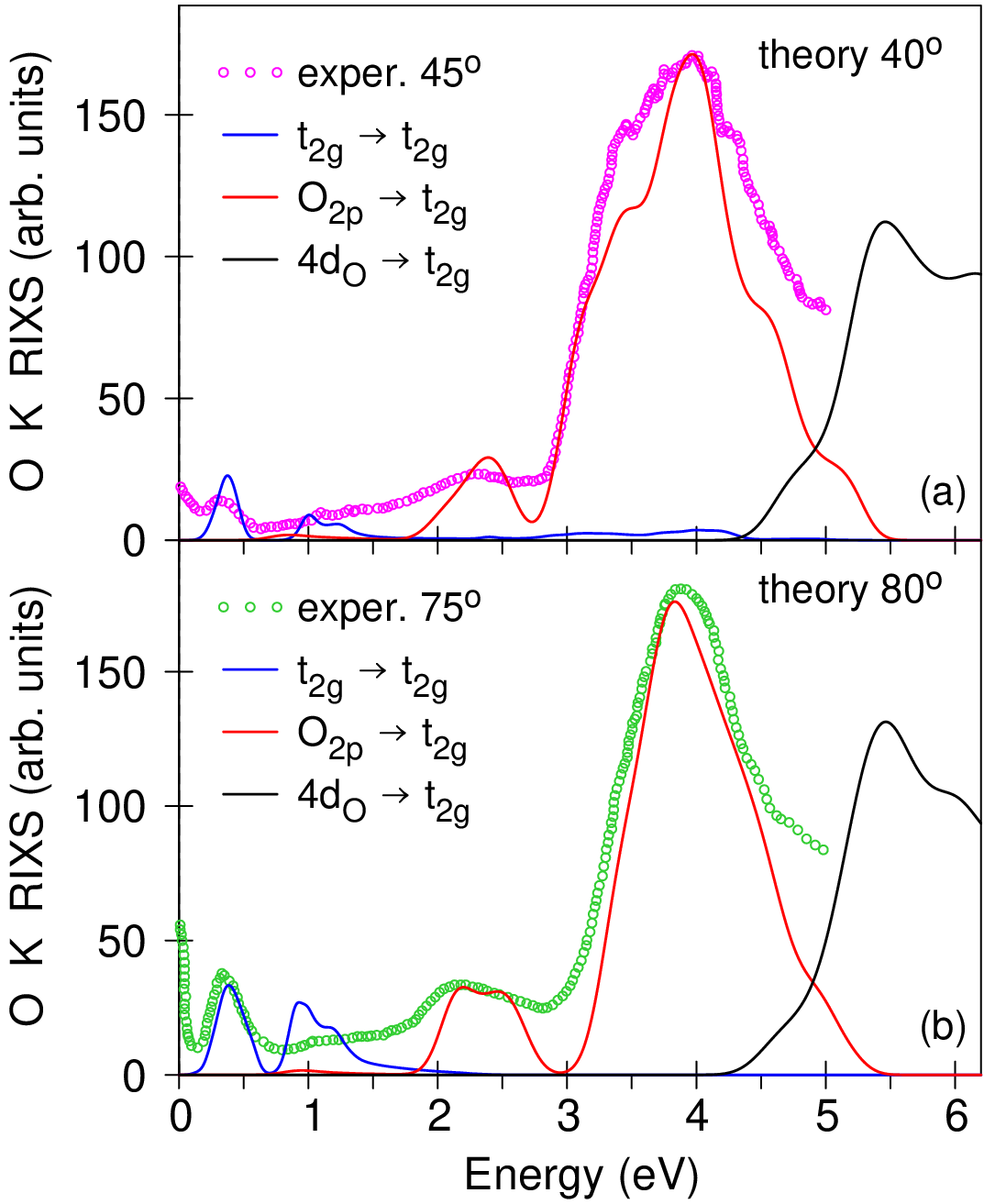}
\end{center}
\caption{\label{rixs_O_CRO}(Color online) The experimentally measured
  O $K$ RIXS spectrum in Ca$_2$RuO$_4$ \cite{DFF+18} (open magenta
  circles) in comparison with the theoretically calculated partial
  contributions from different interband transitions calculated in the
  GGA+SO+$U$ approximation with $U_{\rm{eff}}$ = 0.5 eV. }
\end{figure}

Figure \ref{rixs_O_CRO} shows the theoretically calculated partial
contributions to the O $K$ RIXS spectrum for Ca$_2$RuO$_4$ from different
interband transitions for $\theta$ = 40$^{\circ}$ (the upper panel) and
$\theta$ = 80$^{\circ}$ (the lower panel) in comparison with the experimental data
presented by Das {\it et al.}~\cite{DFF+18}. 

The O $K$ RIXS spectrum consists of three major inelastic excitations:
a double peak at $\le$1.5 eV (blue curves), a major peak between 2 and 5
eV with a low energy shoulder at 2-3 eV (red curves), and a less intensive
structure at 4.5-7 eV (black curves). We found that the first double
peak low energy feature is due to the interband transitions between the
occupied and empty O 2$p$ states that appear as a result of the
strong hybridization between oxygen 2$p$ states with Ru {\tg} LEB and
UEB in the close vicinity of the Fermi level (see
Fig. \ref{PDOS_CRO}), therefore, the oxygen $K$ RIXS spectroscopy can
be used for the estimation of the positions of Ru 4$d$ Hubbard
bands. The major peak between 2 and 5 eV reflects the interband
transitions between the occupied O 2$p$ states and the empty oxygen
states that originate from the hybridization with Ru {\tg}
states. The fine structure between 4.5 and 7.8 eV is due to the
4$d_{\rm{O}}$ $\rightarrow$ {\tg} interband transitions.

The polarization dependence of the RIXS spectrum for the oxygen $K$ edge
presented in Fig.~\ref{rixs_O_CRO} is much weaker compared to the corresponding dependence for the Ru $L_3$ edge. Upon changing from $\theta$ = 40$^{\circ}$ to $\theta$ = 80$^{\circ}$ the two low energy
peaks of the O $K$ RIXS spectrum below 1.5 eV are slightly increased
in intensity and the major peak between 3 and 5 eV becomes more narrow
loosing some fine structures.

\begin{figure}[tbp!]
\begin{center}
\includegraphics[angle=-90, width=0.9\columnwidth]{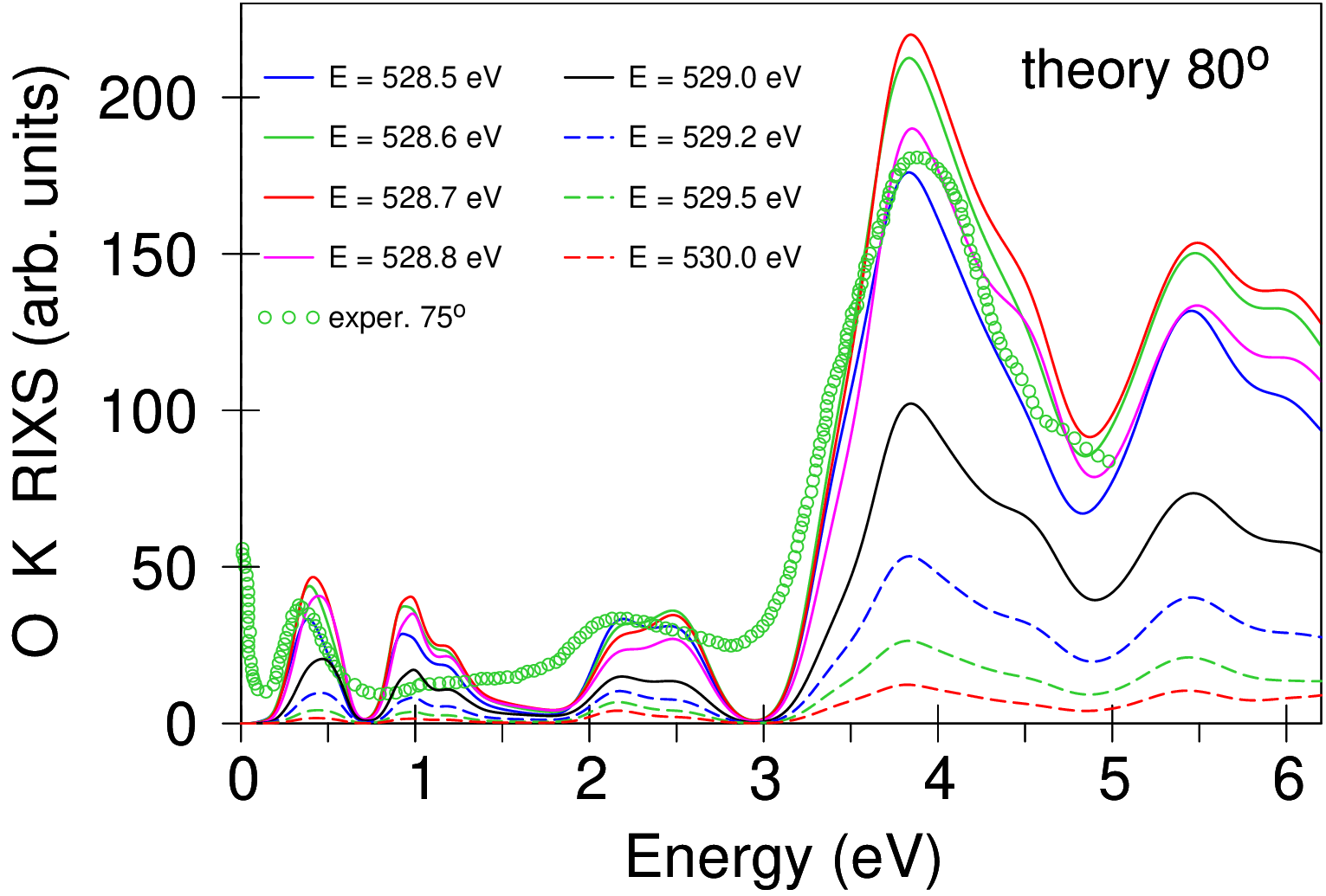}
\end{center}
\caption{\label{rixs_O_Ei_CRO}(Color online) The RIXS spectra as a
  function of incident photon energy calculated at the O $K$ edge in
  Ca$_2$RuO$_4$ for $\theta$ = 80$^{\circ}$ in comparison with the
  experimental measurements for $\theta$ = 75$^{\circ}$
  \cite{DFF+18}. }
\end{figure}

Figure \ref{rixs_O_Ei_CRO} shows the O $K$ RIXS spectra as a function
of incident photon energy calculated in Ca$_2$RuO$_4$ for $\theta$ =
80$^{\circ}$. With increasing the incident photon energy from $E_i$ =
528.5 eV, the peaks increase and then after $E_i$ = 528.7 eV
rapidly decrease and almost vanish for $E_i$ $\geq$ 530 eV.

\section{Key Findings \protect\lowercase{and} Conclusions}

To summarize, we have thoroughly investigated the electronic and magnetic properties of the single-layered perovskite Ca$_2$RuO$_4$ system theoretically, within the fully relativistic spin-polarized Dirac approach. We have performed theoretical modeling of the XES, PES, XAS, XMCD, and RIXS spectra at the Ru $L_{2,3}$ and O $K$ edges, comparing our results with available literature data. Our findings indicate that the intricate interplay of electron correlations, SOC, intersite hoppings, Hund's coupling, as well as the CEF gives rise to a strongly competing ground state in Ca$_2$RuO$_4$.
It is widely accepted that Ca$_2$RuO$_4$ is a Mott insulator, as it remains insulating above the N$\rm{\acute{e}}$el temperature and exhibits Curie–Weiss magnetic susceptibility~\cite{NIM97}. The results of detailed ARPES measurements on Ca$_2$RuO$_4$~\cite{SFM+17} further support the presence of an orbitally differentiated band-Mott insulating ground state.
Ca$_2$RuO$_4$ is reported to undergo a series of phase transitions upon cooling: a metal-to-insulator transition beginning at 357~K, followed by orbital ordering at 260~K, and further by AFM ordering at 110~K. Our band structure calculations reveal that the canted noncollinear AFM ordering AFM$^{\rm{NC}}_{010}$, characterized by an energy gap between the fully occupied \( J_{\rm{eff}} = 3/2 \) and empty \( J_{\rm{eff}} = 1/2 \) states, is energetically favored compared to nonmagnetic, ferromagnetic, or other AFM configurations. This finding highlights the critical role of SOC and electronic correlations in stabilizing the ground state of Ca$_2$RuO$_4$.

SOC splits the {\tg} manifold into a lower $J_{\rm{eff}}$ = 3/2 quartet and
an upper $J_{\rm{eff}}$ = 1/2 doublet in Ca$_2$RuO$_4$. The functions of
the $J_{\rm{eff}}$ = 3/2 quartet are dominated by $d_{3/2}$ states with
some weight of $d_{5/2}$ ones, the $J_{\rm{eff}}$ = 1/2 functions are
almost completely given by linear combinations of $d_{5/2}$
states. From our GGA+SO+$U$ calculations, we obtained the ratio $I_{L_3}/I_{L_2}$, known also as the BR ratio, and employed as a direct probe of the magnitude of SOC, which is found to be 2.46 for Ca$_2$RuO$_4$. Although this value exceeds the statistical ratio in the absence of orbital magnetization, it remains significantly smaller than that observed for the case of iridates with particularly strong SOC, such as Sr$_2$IrO$_4$, where the measured BR approaches a value of 4.1~\cite{HFZ+12}. This implies that SOC effects are essentially weaker in Ca$_2$RuO$_4$ than in these iridate compounds.

The energy gap in Ca$_2$RuO$_4$ opens up only when Hubbard electron-electron correlations are taken into account. However, the effective Hubbard parameter $U^c_{\rm{eff}}$ strongly depends on magnetic ordering. Consequently, Ca$_2$RuO$_4$ exhibits a mixed Slater and Mott character.

We have thoroughly investigated the nature of the RIXS spectra at the Ru $L_3$  and O $K$ edges through theoretical modeling. The best agreement between the calculated and experimentally obtained RIXS spectra was achieved within the GGA+SO+$U$ approximation using the effective Hubbard parameter $U_{\rm{eff}}$ = 0.5 eV.

Ru {\tg} LEB and UEB states have two and one DOS peaks, respectively. The interband transitions between these peaks produce a two major peak structure in the intra-{\tg} excitations. However, the
experiment produces four peaks $A_1$, $A_2$, $A_3$, and $A_4$ at 0.05,
0.32, 0.75, and 1.0 eV below 1.5 eV. The lowest peak $A_1$ at 0.05 eV
can be interpreted as magnetic excitations~\cite{GSK+19} in
consistency with neutron and Raman scattering measurements
~\cite{JKP+17,SCK+17}. There is very strong photon polarization
dependence of the $A_2$ and $A_4$ features. On the other hand, such
dependence was extremely weak for peak $A_3$. 
This may indicate that peak $A_3$, which does not appear in our DFT calculations, has a fundamentally different origin compared to peaks $A_2$ and $A_4$. We propose that peak $A_3$  may be of excitonic nature. A theoretical treatment of magnon and excitonic excitations requires a more advanced approach beyond the one-particle approximation.

The Ru $L_3$ edge RIXS spectrum of Ca$_2$RuO$_4$ reveals several
peaks above the intra-{\tg} excitations at higher energies. Peak
$B$ located between 2.5 eV and 4 eV is mostly due to $\tg \rightarrow
\eg$ transitions with some additional O$_{2p}$ $\rightarrow \tg$ and
O$_{2p}$ $\rightarrow \eg$ transitions. Fine structure $C$ at 5
to 7 eV is due to the 4$d_{\rm{O}}$ $\rightarrow$ {\tg} transitions.
The theoretical calculations are in good agreement with the
experimental data.

We found very strong polarization dependence of the RIXS spectrum at
the Ru $L_3$ edge. The intensity of peak $A_2$ at 0.32 eV is
strongly decreased upon decreasing angle $\theta$ from 0$^{\circ}$
to 90$^{\circ}$. On the other hand, due to the quasi-two dimensional
crystal structure of Ca$_2$RuO$_4$, the momentum dependence of the RIXS
spectrum along the $c$ axis was found to be very small. 

The investigation of the Ru $L_3$ RIXS spectra as a function of
incident photon energy $E_i$ shows that peak $A_2$ is monotonically
decreased with increasing the incident photon energy from $E_1$ =
2838.5 eV, which corresponds to the 4$d_{\tg}$ edge. On the other hand,
the intensity of peak $B$ at 2.5-4 eV, which is derived from the
2$p_{3/2}$ $\rightarrow$ 4$d_{\eg}$ transitions, increases from
$E_1$ to $E_2$ = 2841 eV (which corresponds to the 4$d_{\eg}$ edge)
and then decreases with the further increase of energy.

The O $K$ RIXS spectrum consists of three major inelastic excitations:
a double peak at $\le$1.5 eV, a major peak between 2 and 5 eV with a low
energy shoulder at 2-3 eV, and a less intensive structure at 4.5-7
eV. We found that the first double peak at low energy is due to the
interband transitions between the occupied and empty 2$p$ oxygen states
that appear as a result of the strong hybridization between oxygen
2$p$ states with Ru {\tg} LEB and UEB in the close vicinity of the
Fermi level. The next major peak between 2 and 5 eV reflects the
interband transitions from the occupied O 2$p$ states and the empty
oxygen states that originate from the hybridization with Ru {\tg}
states. The fine structure between 4.5 and 7.8 eV is due to the
4$d_{\rm{O}}$ $\rightarrow$ {\tg} interband transitions. The
polarization dependence of the RIXS spectrum for the oxygen $K$ edge is
much weaker in comparison with the corresponding dependence for the Ru
$L_3$ edge. Upon changing from $\theta$ = 40$^{\circ}$ to $\theta$ =
80$^{\circ}$ the low energy RIXS peaks below 1.5 eV are
slightly increased in intensity and the major RIXS peak between 3 an 5 eV
becomes more narrow loosing some fine structures. 

We found much stronger dependence on the incident photon energy in the
case of the O $K$ RIXS spectrum in comparison with the corresponding
dependence at the Ru $L_3$ edge. With increasing the incident photon
energy from $E_i$ = 528.5 eV, the peaks increase and then after
$E_i$ = 528.7 eV rapidly decrease and almost vanish for
$E_i$ $\geq$ 530 eV.

\section*{Acknowledgments}

We are thankful to Dr. Alexander Yaresko from the Max Planck Institute
FKF in Stuttgart for helpful discussions. This work was supported by the Ministry of Education and Science of Ukraine within the Ukrainian-Austrian Joint Programme of Scientific and Technological Cooperation, 
project "Resonant inelastic x-ray scattering in complex 5d oxides from first principles" (Agrmt. No. M/73-2025, Reg. No. 0125U003444). The authors P.F. Perndorfer, P.A. Buczek and A. Ernst acknowledge the funding by the Fonds zur Förderung der wissenschaftlichen Forschung (FWF) under Grant No. I 5384 / DFG-LAV grant \enquote{SPINELS} BU 4062/1-1.
 

\newcommand{\noopsort}[1]{} \newcommand{\printfirst}[2]{#1}
  \newcommand{\singleletter}[1]{#1} \newcommand{\switchargs}[2]{#2#1}

\end{document}